\documentclass[runningheads]{llncs}

\usepackage{url}

\usepackage{graphicx}
\usepackage{tikz,graphics,color,fullpage,float,epsf,caption,subcaption}
\usepackage{longtable}
\usepackage{caption}
\usepackage{enumitem}
\usepackage{subfiles}

\begin{document}
\sloppy

\title{Towards Data-Enabled Physical Activity Planning: An Exploratory Study of HCP Perspectives On The Integration Of Patient-Generated Health Data}
% add within title if necessary: \thanks{Supported by organization x.} 
\author{Anonymous}
\institute{---}

\author{Pavithren V S Pakianathan\inst{1,2} \and
Hannah McGowan\inst{1} \and
Isabel Höppchen\inst{1} \and
Daniela Wurhofer\inst{1} \and
Gunnar Treff\inst{1} \and
Mahdi Sareban\inst{1} \and
Josef Niebauer\inst{1} \and
Albrecht Schmidt\inst{2} \and
Jan David Smeddinck\inst{1}}
\authorrunning{P.V.S. Pakianathan et al.}
\institute{Ludwig Boltzmann Institute for Digital Health and Prevention, Salzburg, Austria\\
\email{\{pavithren.pakianathan, hannah.mcgowan, isabel.hoeppchen, daniela.wurhofer, gunnar.treff, mahdi.sareban, josef.niebauer, jan.smeddinck}\}@lbg.ac.at \and
LMU Munich, Munich, Germany\\
\email{albrecht.schmidt@ifi.lmu.de}}
\maketitle              % typeset the header of the contribution
\begin{abstract}
Physical activity planning is an essential part of cardiovascular rehabilitation. Through a two-part formative design exploration, we investigated integrating patient-generated health data (PGHD) into clinical workflows supporting shared decision-making (SDM) in physical activity planning. In part one, during a two-week situated study, to reduce risk of working with cardiovascular disease patients, we recruited healthy participants who self-tracked health and physical activity data and attended a physical activity planning session with a healthcare professional (HCP). Subsequently both HCPs and participants were interviewed. In part two, findings from part one were presented to HCPs in a card-sorting workshop to corroborate findings and identify information needs of HCPs alongside patient journeys and clinical workflows. Our outcomes highlight HCP information needs around patient risk factors, vital signs, and adherence to physical activity. Enablers for PGHD integration include adaptive data sense-making, standardization and organizational support for integration. Barriers include lack of time, data quality, trust and liability concerns. Our research highlights implications for designing digital health technologies that support PGHD in physical activity planning during cardiac rehabilitation.
\keywords{digital health, shared decision-making, patient-generated health data, wearables, sensing devices, human-data interaction, prevention, cardiac rehabilitation, cardiovascular disease}
\end{abstract}

\section{Introduction}
Cardiovascular disease (CVD) is the leading cause of death worldwide \cite{noauthor_cardiovascular_nodate}. 
Physical activity planning in cardiovascular rehabilitation helps CVD patients manage their condition and reduce their risk of recurrent cardiac events \cite{winnige2021cardiac}.Traditional physical activity planning \cite{bethellCardiacRehabilitationIt2008,DevelopPhysicalActivity} relies on self-reported routines through questionnaires such as IPAQ (International physical activity questionnaire) \cite{craig_international_2003,craig_international_2017} or information scaffolded through the F.I.T.T (F = frequency, I = intensity, T = time, and T = type) framework developed by the American College of Sports Medicine refers to multiple dimensions of PA \cite{garber_quantity_2011,medicine_acsms_2013}), which often lack objectivity and do not provide HCP with an objective and holistic picture of their patients' activities.

Self-tracking through wearables and mobile health applications enable more comprehensive data collection by the patient. Such multimodal data streams of PGHD are valuable in the interaction between patients and HCPs. PGHD has several benefits such as improving the objectivity of consultations \cite{nittasElectronicPatientGeneratedHealth2019,rutjesBenefitsCostsPatient2017}, enabling more comprehensive patient histories, improving physical activity behaviour \cite{consolvoDesignRequirementsTechnologies2006,tadasUsingPatientGeneratedData2023a}, supporting reflection on health conditions \cite{Ayobi2017} and facilitating patient provider collaboration\cite{chung2016boundary}, thereby improving SDM \cite{lavalleeMHealthPatientGenerated2020g}.

% HCPs are increasingly using telemedicine - the delivery of health care services, where patients and providers are separated by distance \cite{world2019guideline} - to treat chronic patients \cite{guptaHealthDataSciences2024}. 

% Stemming from the quantified-self movement  \cite{gimpelQuantifyingQuantifiedSelf} (cf. Section \ref{sec:pi}) and personal informatics \cite{}, an increasing number of healthy individuals as well as those living with chronic conditions are using self-tracking technologies to monitor their lifestyle and health behaviours.  

Despite its potential, common barriers to PGHD utilization have hindered adoption \cite{west2018common}. For instance, PGHD could (negatively) influence patient-HCP interpersonal dynamics \cite{chung2016boundary,costafigueiredoUsingDataApproach2021l} and increase the burden on HCPs if not well-integrated into clinical workflows \cite{yeImpactElectronicHealth2021a}. Research has shown differences between clinician-initiated tracking, which tends to be more targeted, and patient-initiated tracking \cite{zhu_sharing_2016,tadasUsingPatientGeneratedData2023a}, which is broader and often results in HCP-patient asymmetries \cite{oh_patients_2022}. 

Within the context of physical activity planning in cardiac rehabilitation, although studies have investigated wearable sensor quality \cite{herkert_usefulness_2019} and their effectiveness in increasing physical activity \cite{ashur_wearable_2021}, there is lack of nuanced data about HCPs perspective on the challenges and opportunities for integrating PGHD for PA planning in cardiovascular rehabilitation. The integration of PGHD from consumer wearables requires thinking beyond traditional clinical boundaries emphasizing the need for a deeper understanding of implication for patients and HCPs \cite{tadas_user-centred_2022}. Human-computer interaction (HCI) research takes a holistic approach to designing digital health technologies by emphasizing processes over stand-alone technologies, however, main-streaming digital health technologies requires buy-in from HCPs \cite{hoppchenBeMeStay2024}. Understanding the needs and routines of these key stakeholders along the patient data journey and clinical workflows - patients and HCPs - in and out of clinical settings and the multi-level factors within an organizational context are crucial for designing digital health tools which facilitate the integration of PGHD in physical activity planning \cite{nunes2015self,duran2023applying,andersenAligningConcernsTelecare2019}. 

There is lack of nuanced data about HCPs perspective on the challenges and opportunities for integrating PGHD for PA planning in cardiovascular rehabilitation.

This paper focuses on HCP perspectives on how PGHD from self-tracking technologies can be integrated into clinical workflows and patient data journeys to support physical activity planning in cardiovascular rehabilitation. Due to the exploratory nature of our study and to reduce any (psychological) risks for vulnerable CVD patients such as anxiety \cite{rosman_when_2020,varma2024promises}, we decided to recruit healthy subjects. The aim of our formative design exploration is to understand how it can be integrated into clinical workflows and patient journeys to improve physical activity planning during cardiac rehabilitation of CVD patients. \textbf{We extend PGHD and SDM literature by framing physical activity planning in cardiac rehabilitation as a multi-level data work problem, combining in-situ observations with structured co-design to bridge HCI design and implementation science.} By doing so, we offer actionable preliminary design implications for supporting PGHD integration, providing insights for the pathway for data-enabled physical activity planning sessions.

\subsection{Research Questions}
\label{sec:rq}
This paper is based on the following research questions (RQs): 
\begin{enumerate}[label={}]
    \item RQ1: What are HCP perspectives on the challenges and opportunities for integrating PGHD for physical activity planning in cardiovascular rehabilitation?
    \item RQ2: What design considerations are relevant for real-world implementations of data-enabled physical activity planning pathways in the context of cardiovascular rehabilitation?
\end{enumerate}
\section{Background and Related Work}

\subsection{Physical Activity Planning in Cardiovascular Rehabilitation}

During physical activity planning sessions, a key component of cardiovascular rehabilitation, a patient and a HCP undergo a process of SDM to construct a personalised physical activity plan. This is similar to other chronic conditions like obesity prevention or management or marathon preparation (where coaches rather than HCPs co-create journey with clients).
% The process of SDM is essential for patient-centred care and has been related to improved treatment outcomes, higher patient satisfaction levels \cite{kuosmanen2021patient} and treatment adherence \cite{homketowska2023adherence}. 
In the context of physical activity planning, a HCP first needs to understand the patient's risk profile, physical activity levels, preferences, social context and other potential barriers and enablers before engaging in a SDM process.  \cite{hoppchenTargetingBehavioralFactors2024}. Afterwards, in the context of the patient journey of cardiac rehabilitation, regular follow-up consultations occur to inquire and check if patients are adhering to their recommended lifestyle behaviours. 

As care models evolve, they necessitate shifts in clinical flows and patient journeys. The integration of PGHD - such as physical activity levels, heart rate, weight, blood pressure - from consumer wearables into physical activity consultations has the potential to improve SDM by allowing HCPs to have a comprehensive understanding of their patients' physical activity behaviour and health, enabling personalized recommendations \cite{suWhatYourEnvisioned2022}. To effectively understand how PGHD integration can be implemented, it is necessary to align with HCPs' nuanced clinical routines while ensuring acceptability and sustainability. This would in turn, minimize potential burdens on HCP \cite{serbanJustSeeNumbers2023,greenhalghAdoptionNewFramework2017}.

\subsection{Personal Informatics Across the Chronic Care Continuum}
\label{sec:pi}

Self-tracking technologies (e.g., smartwatches, wristbands, mobile applications) are increasingly being used by both healthy individuals and those with chronic conditions to monitor and reflect on their health and behaviors \cite{liStagebasedModelPersonal2010,choeCharacterizingVisualizationInsights2015,mishraSupportingCollaborativeHealth2018}. They enable the collection of various physiological and health data (blood oxygen levels, electrocardiogram (ECG), steps, etc.) for consumers. When utilized in a health context such self-tracking or self-reported data is referred to as PGHD. Overall, self-tracking technologies have enabled a permeation of self-tracking abilities in the general public \cite{gimpelQuantifyingQuantifiedSelf} with findings showing overall positive outcomes, such as increased motivation to achieve personal health goals \cite{karanam2014motivational} and in the context of cardiac rehabilitation PGHD has shown to significantly increase physical activity behaviour among CVD patients \cite{ashur_wearable_2021}. 

Despite the positive aspects of self-tracking, individuals may also face barriers to adoption (e.g. lack of time, motivation, forgetfulness, self-criticism and poor usability) \cite{liStagebasedModelPersonal2010,tadas_user-centred_2022}, which in turn could lead to non-adoption or declined engagement. To address these barriers and improve engagement, researchers have proposed frameworks and models that align technology use with individual needs and behavior. Li \emph{et al.} \cite{liStagebasedModelPersonal2010} designed a \emph{staged-based model of personal informatics}, which aimed to better integrate digital health technologies with an individual's lifestyle emphasizing the need for personal informatics systems to be ``designed in a holistic manner across the stages'' of self-tracking. 

PGHD are gaining acceptance in healthcare settings and even among national level policy recommendations \cite{HttpsWwwGov2014}. National infrastructures, such as Denmark’s wearable data system \cite{danish2018coherent}, and platforms for ecological / situated and longer-term data collection, such as Radarbase \cite{radar2019} and MORE \cite{v_s_pakianathan_multi-stakeholder_2023} support PGHD integration in healthcare \cite{kumar_mobile_2021}.However, the successful implementation of personalized digital care pathways depends on several factors, including stakeholder engagement, perceived system usefulness, organizational readiness, usability \cite{heijsters2022stakeholders} and reimbursement models \cite{sareban2025op}. For effective integration of PGHD into clinical workflows, the perspectives of patients and HCPs should  be carefully investigated \cite{oh_patients_2022} particularly in the context of physical activity planning.

\subsection{Data Work and Collaboration in Healthcare}
Integrating PGHD into clinical workflows can personalize care, facilitate discussions and allow HCPs to understand patients \cite{chungMoreTelemonitoringHealth2015} . However, adoption beyond tele-consultations is currently low due to disruption to HCP's workflows and adding ``data work'' \cite{fiske2019health} (e.g., managing and analyzing data) to the already busy HCPs \cite{pine2018data,ding2020effects,davis2014systematic,aranki2016real,bossen2019data,serbanJustSeeNumbers2023}.

Effective use of PGHD in chronic care requires HCP interpretation. However information asymmetries (between tracked and shared data) due to privacy, patient-HCP relationships, and perceived relevance \cite{oh_patients_2022,chung2016boundary} create information gaps for HCPs \cite{oh_patients_2022}, hindering the potential of PGHD. Research has shown that patients and HCPs often lack effective means for working with PGHD during consultations \cite{schroeder2017supporting}. To further investigate such challenges, using Li \emph{et al.}'s 5-stage model of personal informatics \cite{liStagebasedModelPersonal2010} as a guide, Chung \emph{et al.} \cite{chung2016boundary} used the concept of boundary-negotiating artifacts and offered design recommendations on how HCPs and patients can better collaborate with data by setting goals on tracking, sharing and reviewing. Given the highly collaborative and social nature of data work in clinical context, Chung \emph{et al.} \cite{chung2016boundary} and Mishra \emph{et al.} \cite{mishraSupportingCollaborativeHealth2018} emphasised that the stage-based model \cite{liStagebasedModelPersonal2010} has to be extended to consider patient and HCP needs during \textit{collaborative tracking}.
It is necessary to recognize that the needs and workflows of HCPs are not monolithic. As noted by Serban \textit{et al. } effective PGHD integration requires role-specific data representations and analyses. Ultimately, effective and efficient data sensemaking is crucial for meaningful data work for different HCPs.

Prior research also emphasised the importance of supporting collaborative PGHD interpretation to address the needs of both HCPs and patients. Schroeder et al. \cite{schroeder2017supporting} found that interactive visualizations of symptom journals fostered mutual understanding and trust with both groups desiring pre-appointment access. HCPs specifically needed quick pre-consultation interpretation for better preparation \cite{chungMoreTelemonitoringHealth2015} and both HCPs and patients emphasized the need for collaborative assessment of data reliability \cite{cernaChangingCategoricalWork2020}. While prior works have provided promising design recommendations, there is still a significant adoption gap of PGHD due to the barriers HCPs face in uptaking PGHD in cardiovascular rehabilitation.

\subsection{Research gap and motivation}

Physical activity planning in cardiac rehabilitation involves multiple stakeholders and extends beyond acute clinical settings \cite{duran2023applying,heijsters2022stakeholders}. Research on the integration of PGHD from consumer wearables exists across a range of chronic conditions \cite{serbanJustSeeNumbers2023,aranki2016real,haase2023data,tadasUsingPatientGeneratedData2023a}. While these studies provide valuable insights into HCP needs and data-based tasks in clinical contexts, the specific challenges and opportunities for integrating PGHD in the highly specialized domain of physical activity planning in outpatient cardiovascular rehabilitation are not yet fully explored. Unlike other chronic conditions, physical activity planning in cardiac rehab involves a unique and high-risk patient population. These individuals often have specific physiological traits, are on complex medication regimens, and require highly tailored exercise prescriptions to manage risk and promote recovery, making the information needs of HCPs unique. Given the potential data overload for clinicians and patients during shared decision making, it is necessary to understand the ``data-based tasks'' HCPs perform \cite{scholich_augmenting_2024} during physical activity planning consultations while integrating PGHD from consumer wearables.

Implementing digital health technologies is often demanding, with adoption lagging behind promising trial results \cite{bauer2020implementation,damschroder2022updated,heijsters2022stakeholders}. Furthermore, balancing patient needs with HCP workflows and organizational processes is challenging  \cite{greenhalghAdoptionNewFramework2017}, worse yet with collaborative interpretation \cite{mishraSupportingCollaborativeHealth2018,chung2016boundary}, where the interpretation of complex, multi-modal data must be negotiated between HCPs and patients with varying information needs and perspectives. Implementation science emphasizes the importance of contextual understanding for successful and sustainable adoption \cite{hoppchenBeMeStay2024,damschroder2022updated,damschroderFosteringImplementationHealth2009,glasgow2014implementation}. HCPs spend significant time interacting with potentially burdensome electronic health records \cite{sinsky_allocation_2016}, contributing to burnout and negatively impacting patient interactions and increasing medical errors \cite{budd_burnout_2023,forde-johnston_integrative_2023,tawfik_physician_2018}. So it is crucial to understand and address their needs concerning novel digital health technologies already in the design process. With the consideration that digital health innovations must be designed with sensitivity to the context \cite{hoppchenBeMeStay2024}, we adopt the NASSS framework \cite{greenhalghAdoptionNewFramework2017} to contextualize our findings.
\section{Method}
\label{sec:method}
\subsection{Investigating the Context: Stakeholders and Technology}

\begin{figure}[htbp]
\centering
\includegraphics[width=0.95\linewidth, alt={Flow diagram illustrating the two-part mixed-methods study process. Part 1 involved six participants using a device for two weeks to collect data, attending a 30-minute physical activity planning session, and participating in a 15–30 minute individual interview with two healthcare professionals. Data were analyzed before Part 2, in which the two healthcare professionals took part in a 1.5-hour card sorting workshop.}]{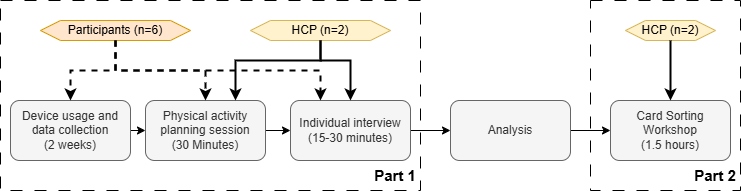}
\caption{Flow, key elements and methods for the two-part study process.}
\label{fig:studyFlow}
\end{figure}

We employed a mixed-methods, formative design exploration (Figure \ref{fig:studyFlow}) consisting of two parts. Part 1 was a situated study with HCPs and healthy participants. This was a calculated decision to reduce burden and potential (psychological) risks (even if minor) for CVD patients in acute clinical or healthcare settings \cite{rosman_when_2020,varma2024promises} at this formative research stage. Part 2 was a workshop with HCPs, focused on designing data enabled workflows to better integrate PGHD physical activity planning in cardiovascular rehabilitation. 

% \begin{figure}
%     \centering
%     \includegraphics[width=0.9\linewidth]{images/Study_Flow.png}
%     \caption{Flow, key elements and methods for the two-part study process.}
%     \label{fig:studyFlow}
% \end{figure}

The participants were informed about the objective of the study and that they would be engaging in a physical activity planning session after the data collection period. We recruited smartphone users and those who are comfortable with using wearables for self-tracking for the purpose of improving physical activity behaviour. In the following sections, the participants who brought their self-tracked data to consultations would be referred to as participants while healthcare professionals are referred to as HCPs. Ethics approval was obtained from the institution for conducting the studies. Participant information sheet was shared with the participants and the HCP and participants gave written consent for taking part in the study, voice recording during consultations with HCPs and interviews and taking non-personally identifiable photographs during the consultations. Health data was not collected from the participants by the researchers and purely used to support the physcial activity planning session.

Part 1 offered insights into the contextual factors which influence how HCPs collaborate with participants to interact with PGHD for physical activity planning. Part 2 included a card-sorting workshop with HCPs to design a clinical workflow supporting effective data sensemaking in data-enabled physical activity planning. Ethics were approved by the Ludwig Boltzmann Gesellschaft research ethics committee (reference 003-2023).

\subsection{Part 1}
\subsubsection{Recruitment and Study Setup}
Recruitment was facilitated via publicity on a local Facebook page and through colleagues' networks. Participants received a compensation of 50 EUR. HCPs were recruited via the local hospital in Salzburg who are affiliated with our research institute and had experience in physical activity planning with CVD patients. 

Participants used a Withings smartwatch, weighing scale and a blood pressure monitor \footnote{Smartwatch (ScanWatch), Body Composition Scale (BodyScan and BodyCardio) and Blood Pressure Monitor (BPM Connect)} - over two weeks before an HCP consultation. The devices were selected after an internal usability and feasibility assessment with colleagues who have experience in cardiac rehabilitation and based on offering an integrated solution with a suite of devices and a single companion application with a health-oriented focus, including an existing PDF report generation functionality. The purpose of using this set of tools was to have a setup which facilitates the understanding of integrating self-tracked data from consumer wearables in a realistic way. We collected participant's physical activity using the Rapid Assessment of Physical Activity (RAPA) questionnaire \cite{topolskiPeerReviewedRapid2006} to characterise them but these details were not shared with the HCPs. They received detailed instructions on tracking and were encourage to setup and troubleshoot independently with minimal researcher involvement.

% \begin{figure}[h]
%     \centering
%     \begin{subfigure}{0.3\textwidth}
%         \centering
%         \includegraphics[width=\textwidth]{images/contextualInquiry/P2.JPG}
%         \caption{P2 consultation with HCP2}
%     \end{subfigure}
%     \begin{subfigure}{0.3\textwidth}
%         \centering
%         \includegraphics[width=\textwidth]{images/contextualInquiry/P4.JPG}
%         \caption{P4 consultation with HCP1} 
%     \end{subfigure}
%     \begin{subfigure}{0.3\textwidth}
%         \centering
%         \includegraphics[width=\textwidth]{images/contextualInquiry/P6a.JPG}
%         \caption{P6 consultation with HCP2} 
%     \end{subfigure}
%     \caption{Impressions from observed consultation sessions between participants and HCPs}
%     \label{fig:consultation}
% \end{figure}

\begin{figure}[h]
    \centering
    \begin{subfigure}{0.3\textwidth}
        \centering
        \includegraphics[width=\textwidth, alt={Photograph of a consultation where a participant and a healthcare professional view health data on a smartphone.}]{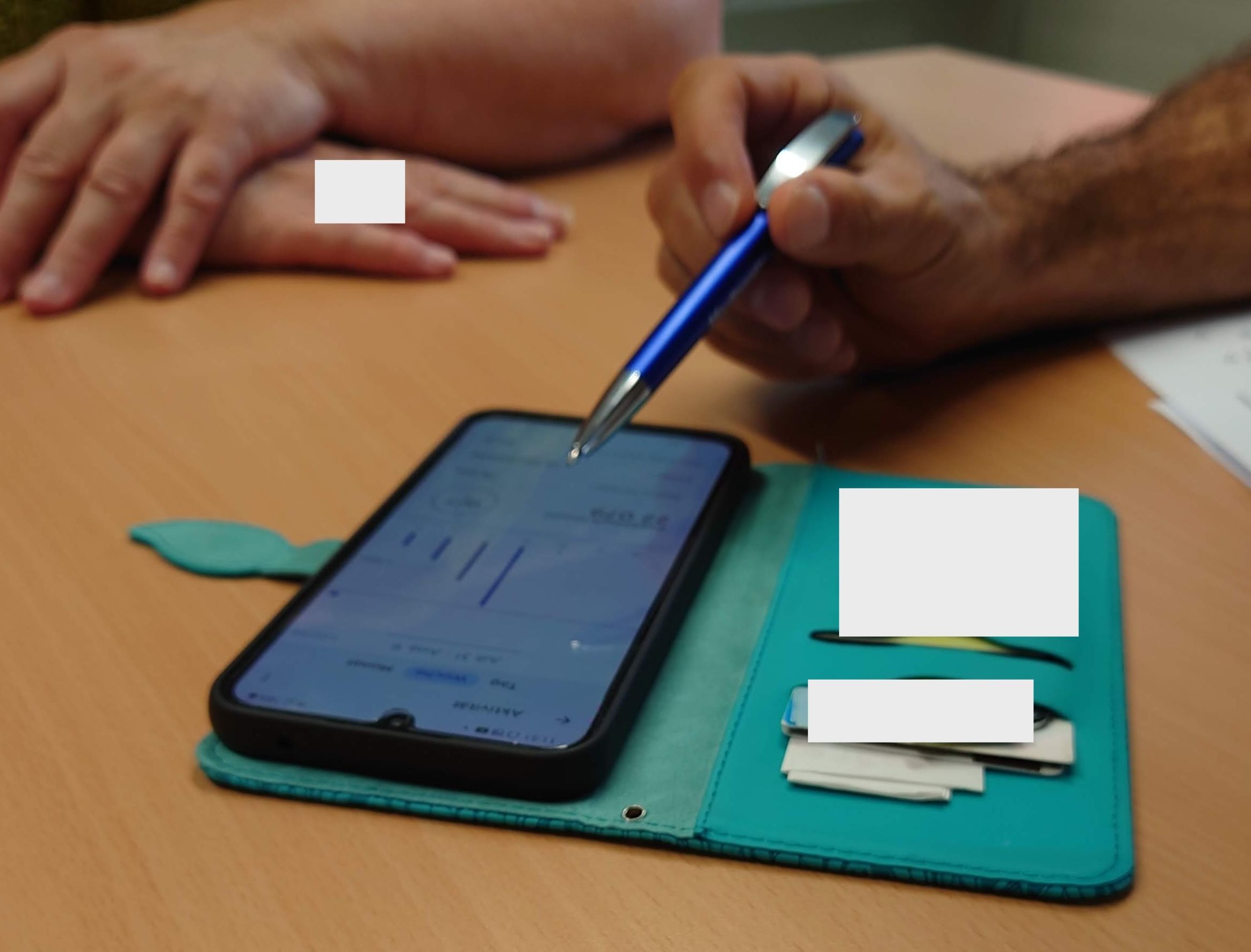}
        \caption{P2 consultation with HCP2}
    \end{subfigure}
    \begin{subfigure}{0.3\textwidth}
        \centering
        \includegraphics[width=\textwidth, alt={Photograph of a consultation where a participant and a healthcare professional review a printed sheet with recorded measurements.}]{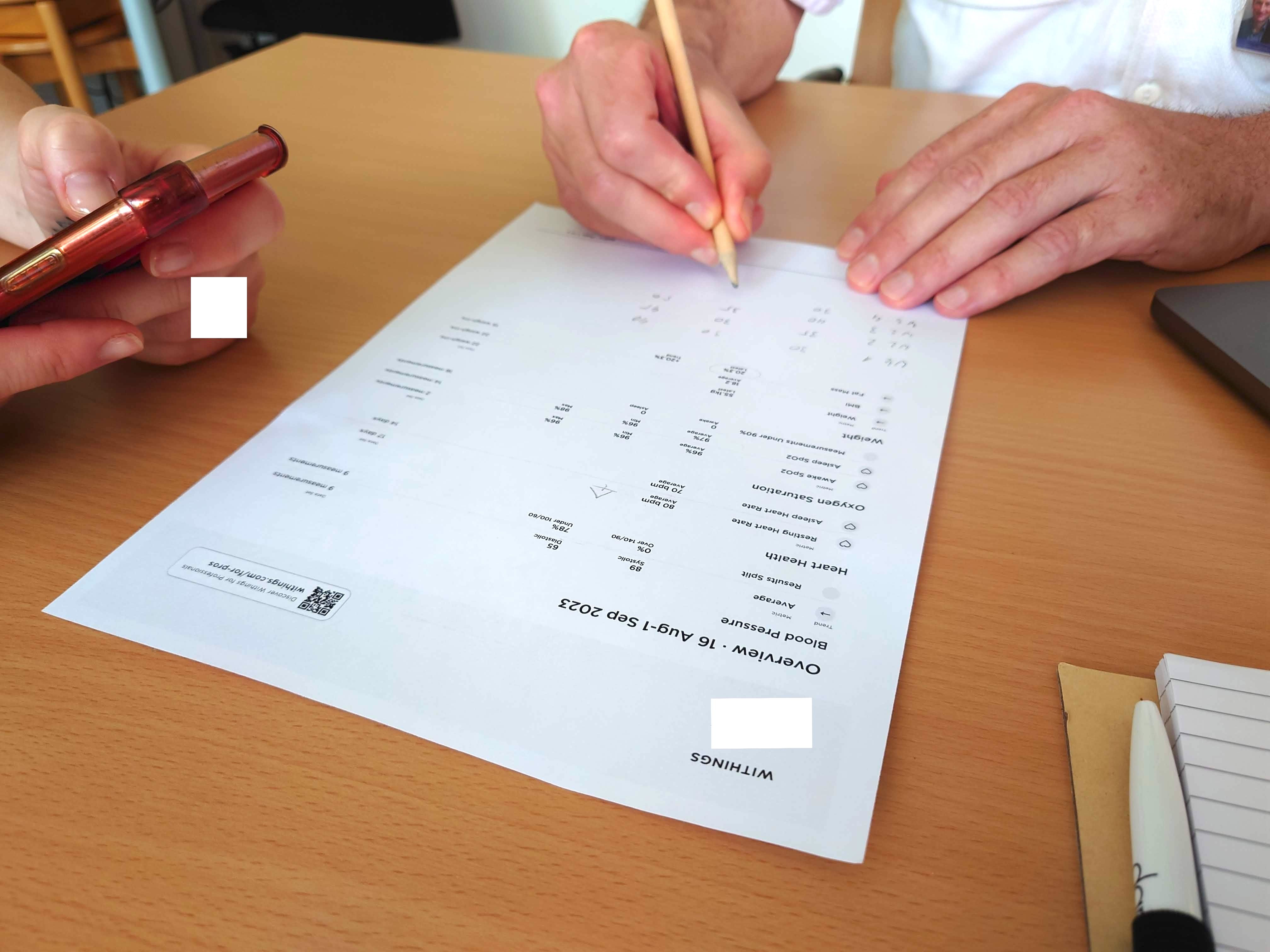}
        \caption{P4 consultation with HCP1} 
    \end{subfigure}
    \begin{subfigure}{0.3\textwidth}
        \centering
        \includegraphics[width=\textwidth, alt={Photograph of a consultation showing hands with a pen while discussing results on a printed form.}]{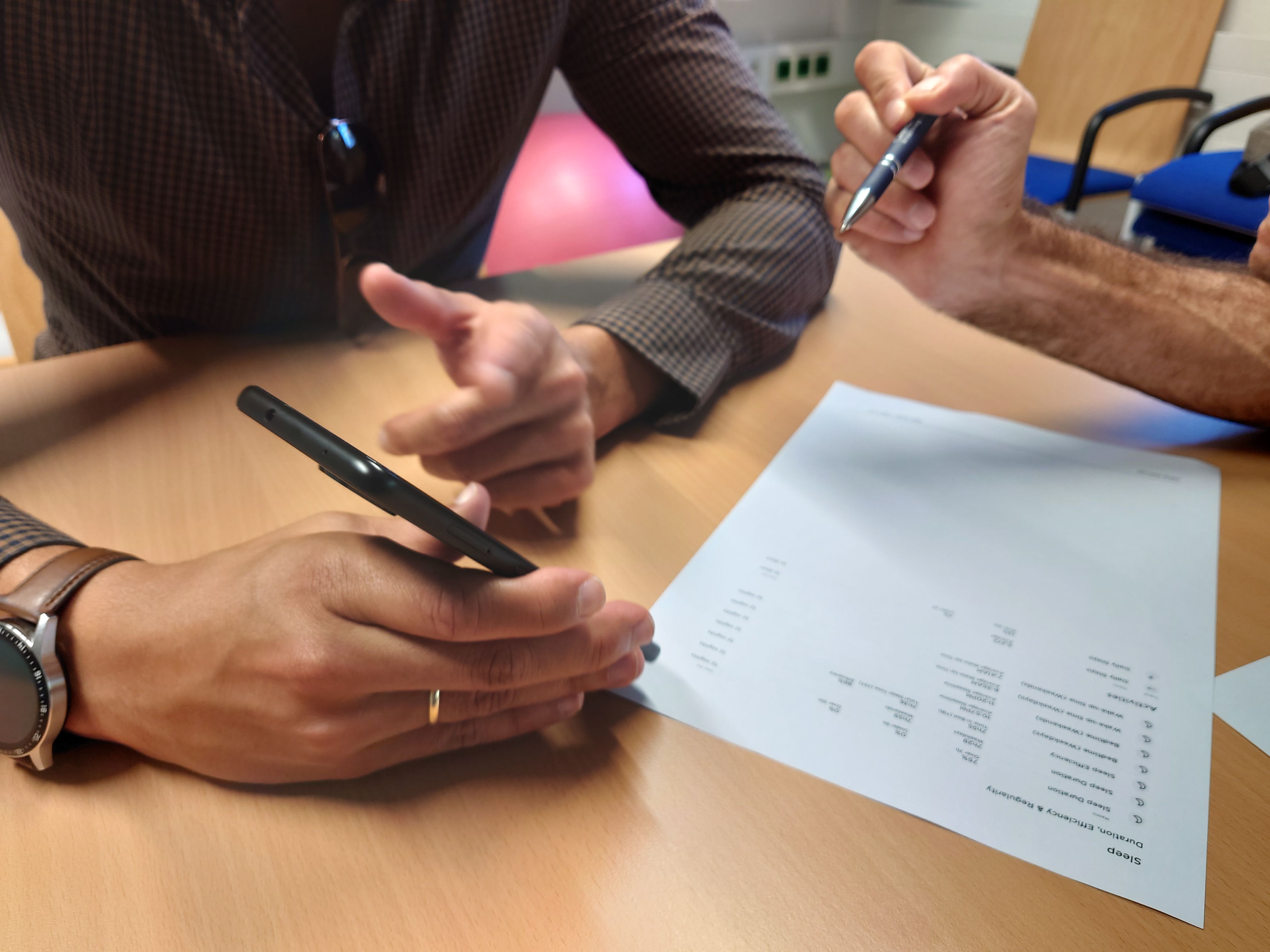}
        \caption{P6 consultation with HCP2} 
    \end{subfigure}
    \caption{Impressions from observed consultation sessions between participants and HCPs.}
    \label{fig:consultation}
\end{figure}

While the smartwatches were used for the entire two weeks, the blood pressure monitor and weighing scale were provided to participants in a permutated order for 7 days of consecutive use each. Participants were instructed to wear the smartwatch while sleeping, measure at least one activity of at least 10 minutes during a physical activity workout, capture electrocardiogram (ECG) and oxygen saturation (SpO2) on day 7 and day 14 using the smartwatch, take daily blood pressure measurements using the smart blood pressure monitor, as well as daily weight and body composition measurements (including body fat, weight, and body mass index) and pulse wave velocity (arterial stiffness measurement to assess cardiovascular disease risk) using the smart weighing scale. A tracking period of at least 7 days was chosen to obtain sufficient data to facilitate the consultation. The data points were selected after consultation with colleagues who specialize in cardiac rehabilitation and to make the study setup as realistic as possible especially because in a realistic scenario, patients could be tracking multiple variables (e.g. having obesity, high BP).

Physical activity planning sessions (c.f. Figure \ref{fig:consultation}) were conducted with two HCP one cardiologist specialized in cardiac rehabilitation and one sports scientist with a focus on exercise physiology and rehabilitation science) with more than 15 years of practical experience each, including regularly engaging in physical activity planning.

\subsubsection{Observation of physical activity planning sessions}
For the physical activity planning session, participants brought a printed summary report available from the Withings application, along with the mobile application. The summary report contained an overview of the physical activity history and vital sign measurements during the two weeks of self-tracking. As patients have already been known to bring printed or logged artifacts for consultations, this step was deliberately added to see how such a printed report from a wearable provider could influence the discussion. Before the session, both HCPs and participants were encouraged to use the application to support shared decision-making. To ensure grounded action, the HCP entered a proposed physical activity plan into a physical activity scheduling application \cite{gartner2025digitally}. A researcher observed the consultation session and took field notes with a focus on interpersonal dynamics, as well as data and technology utilization. The consultation was audio-recorded. Photos to illustrate the context were taken (c.f. Figure \ref{fig:consultation}) with care not to include any personally identifiable information.

\subsubsection{Interviews with Participants and HCPs}
The HCPs attended a thirty-minute interview with the researcher to share feedback about the experience working with participants during the consultations and their perspectives on the use of PGHD for consultations. Fifteen-minute interviews were also conducted with participants to understand 1) how they felt about the session, especially the experience of sharing the PGHD with the HCP, 2) their experience with the usage of the application and data collection using the devices, and 3) how the data they brought influenced the consultation.

 % For additional member checking, both HCP are also involved as co-authors of this work.
\subsubsection{Analysis}
All interviews and observations were transcribed locally using OpenAI Whisper \footnote{https://openai.com/index/whisper/} and subsequently verified. The data was iteratively coded by words or sentences based on thematic analysis by using Braun and Clarke's thematic analysis method \cite{braun2012thematic} by the first and third author with experience in digital and public health. Firstly, both  familiarized themselves with interview transcripts from one participant and one HCP and generated open codes. Codes were collaboratively checked and code conflicts were discussed till a consensus. Subsequently, the first author continued to code the remainder of the data using MaxQDA \footnote{https://www.maxqda.com/}. The first and third authors then discussed the final codes and derived the themes. 

\subsection{Part 2}

\subsubsection{Workshop (WS) with HCPs}
Building on the findings from Part 1, we held 1.5-hour workshops with HCPs from the same facility to explore how to better integrate PGHD into clinical workflows and organizational contexts. HCPs with a focus on exercise physiology were recruited within the network of the rehabilitation department of the local hospital in Salzburg. A pilot workshop involving 4 medical doctors, each with over 10 years of prior medical practice experience but no current clinical practice, informed the development of the workshop content by streamlining the activities to fit within the allocated 1.5 hours. However, this pilot was excluded from analysis due to a focus shift towards practical application of physical activity planning. 

The final workshops involved two sport scientists specializing in exercise physiology and experienced in physical activity planning in cardiac rehabilitation with one to four years of experience. While different HCPs took part in Part 2 to broaden perspectives and due to availability, they were asked to review and confirm the Part 1 findings to ensure continuity between the two stages.

We started the workshop with an introduction of the topic and a presentation of the summary of findings from Part 1. We asked HCPs to share their perspectives on these findings in a printed questionnaire. Next, we ran a card based activity inspired by the works by Serban \emph{et al.} \cite{serbanJustSeeNumbers2023}. Card-based workshops have shown to initiate idea generation during design processes \cite{InspirationCardWorkshops,royCardbasedDesignTools2019,sandersGenerativeToolsCodesigning2000}. Since we were investigating in the domain of data, card-based workshops have been shown to help materialize data into tangible card formats \cite{yiminlimDatastormingCraftingData2021}. We focused on the needs of HCPs who conduct physical activity planning sessions and their views on how patients might reflect on their self-tracked data during or also before consultations through three key scenarios:
\begin{enumerate}
        \item \textbf{Scenario 1}: How can patients be better informed about themselves and their physical activity levels \textbf{before} physical activity planning sessions? i.e. How can patients reflect on their PGHD before a physical activity planning session?
        \item \textbf{Scenario 2}: What information would HCPs like to know about their patients before meeting them for a physical activity planning session and how can PGHD address them?
        \item \textbf{Scenario 3}: How can HCPs and patients better collaborate with PGHD \textbf{during} physical activity planning sessions?
        \begin{enumerate}
            \item What information is required/useful for the patient to leave with \textbf{after} the physical activity planning session?
            \item What are the information needs of both stakeholders for better SDM?
        \end{enumerate}
\end{enumerate}

\begin{figure}
\centering
    \begin{minipage}{0.51\textwidth}
        \centering
        \includegraphics[width=\linewidth, alt={Diagram showing artifacts for a card sorting activity. The left column lists four categories: Data, Time, Analytics, and Visualization. Each category has an associated card: Steps and Context for Data, Week for Time, Max/Min for Analytics, and Bar Chart for Visualization. On the right, an example objective is given: “I want to know what the patient did when they had the highest and lowest step counts last week.” Below, a photograph displays printed cards labeled Context, Steps, Week, Bar chart, and Find Extremum (Max/Min) arranged on a table.}]{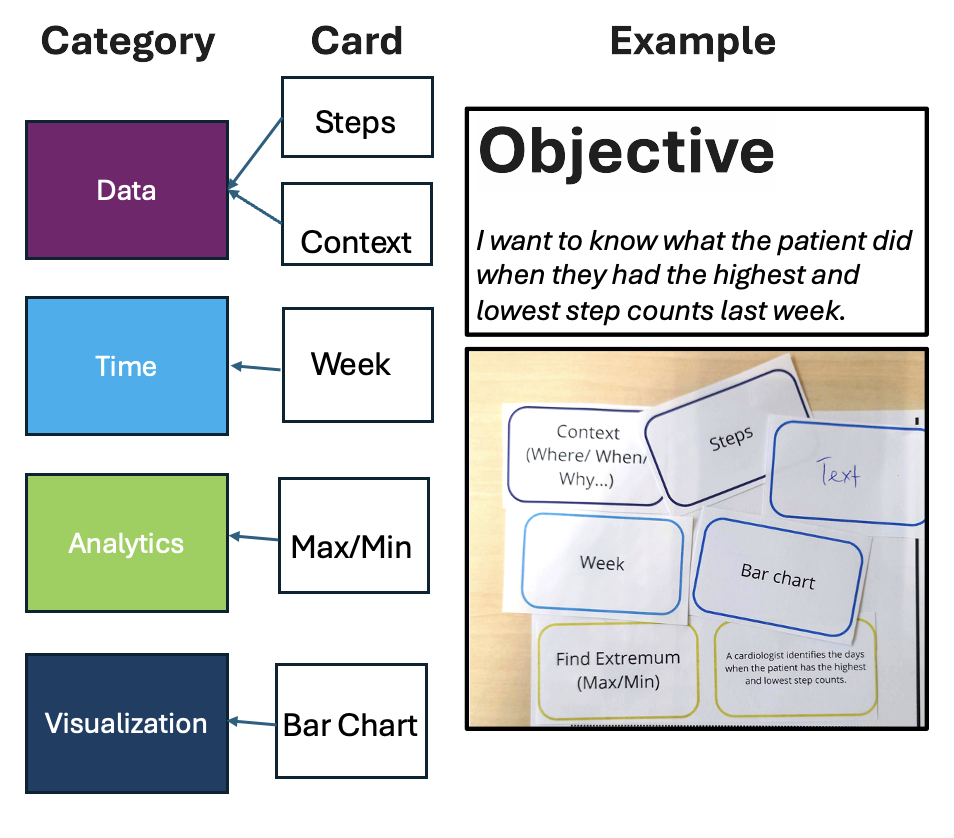}
    \end{minipage} 
\caption{Artifacts for the card sorting activity with example scenario.}
\label{fig:eg_card}
\end{figure}

The three scenarios were derived from feedback from the HCPs in \textbf{Part 1} and also from the 10-stage workflow model for integrating PGHD \cite{VSPakianathan2024} with a specific focus on initial consultations.
Cards covered four categories derived from Part 1 findings: \textbf{1)} \textbf{Data} (Figure \ref{fig:data_card}), \textbf{2)} \textbf{Analysis} (Figure \ref{fig:analysis_card}), \textbf{3)} \textbf{Time} (Figure \ref{fig:time_card}) and \textbf{4)} \textbf{Visualization} (Figure \ref{fig:viz_card}). The \textbf{Data} cards were based on the information needs identified in Part 1, through consultation with a researcher specializing in health informatics in the field of cardiac care. \textbf{Analysis} cards were inspired by the seven tasks guided by the \textit{Visual Information Seeking Mantra} by Shneiderman \cite{shneidermanEyesHaveIt1996}. They had a brief example description printed on the back side for easier comprehensibility (Figure \ref{fig:analysis_card}). Empty cards were also included in each category to allow clinicians to add missing points.
HCPs were asked to group the printed cards under the various categories to create thematic groupings reflecting the information that end users are expected to gain. E.g. Patient can check the maximum/minimum (\textbf{analysis}) step counts (\textbf{data}) over the past week (\textbf{time frame}) as a bar chart (\textbf{visualization}) (Figure \ref{fig:eg_card}). They were also asked to annotate their ideas to aid analysis. After the card-sorting activity, we conducted a MoSCoW prioritization activity (Figure \ref{fig:moscow}) \cite{clegg1994case} to sort the data points into M - Must have, S - Should have, C - Could have, W - Won’t have, to prioritize the data which should be integrated into a data-driven clinical workflow. 
Lastly, the HCPs were asked to brainstorm on common barriers, enablers and other perspectives to the integration of PGHD in clinical workflows. They used post-its notes to write their ideas and paste on a printed sheet.

% \begin{figure}
% \centering
%     \begin{minipage}{0.51\textwidth}
%         \centering
%         \includegraphics[width=\linewidth]{images/EG_Card_Workshop.png}
%     \end{minipage} 
% \caption{Artifacts for the card sorting activity with example scenario}
% \label{fig:eg_card}
% \end{figure}

\subsubsection{Analysis}
Observation notes were taken by the co-facilitator and voice was recorded. The cards and annotations were analysed using Mayring’s structured approach to systematically categorize textual data \cite{mayringQualitativeContentAnalysis2004} by the first and second author. Audio recordings were transcribed using OpenAI Whisper locally and subsequently verified by one researcher who went through the entire transcripts. Observation notes and open-ended responses to the questionnaires were analysed using inductive thematic analysis \cite{braun2012thematic}. Throughout the thematic analysis process, coding was conducted independently and iteratively refined through discussion. Coding discrepancies were resolved through discussion until a consensus was reached.
\section{Findings}
Our study explores potential avenues for design when self-tracked PGHD from consumer wearables is integrated into physical activity planning in cardiac rehabilitation. Our analysis addressed two key research questions (c.f. Section \ref{sec:rq}) - \textbf{RQ1 (challenges and opportunities, RQ2 (design considerations)} - to gain a closer understanding of the context around PGHD integration in physical activity planning workflows in cardiac rehabilitation.  Our 2-part study provided a deep contextual understanding of how PGHD supported collaboration between HCPs and participants during physical activity planning. HCPs also collaborated and contributed their views on the direct practical utilization of PGHD, as well as how PGHD can be integrated into clinical pathways. 

Findings reveal that although PGHD offered better objectivity and personalization of consultations, several data analysis workflow adaptations are required and organizational and systemic barriers such as time constraints, data trust and liability should be addressed by practitioners for effective implementation. 

\begin{figure}
    \centering
    \includegraphics[width=0.9\linewidth, alt={Diagram showing healthcare professionals’ perspectives on barriers and enablers of PGHD across levels of the NASSS Framework. Organizational level: Integration and Interoperability (F5). Adopter level—Pre-Consultation: Enabler is patient education on health data literacy (F8). During Consultation: Enablers include glanceable overview of patient (F5), standardized tools for data interaction (F5), and personalized care through objectivity and patient education (F1). Challenges include unnecessary data being distracting (F4), liability concerns (F6), lack of time (F2), missing information about quality and reliability (F4), and patients needing co-interpretation (F3). Information needs specific to CVD patients (F7) include risk factor, physical activity adherence, and vital sign.}]{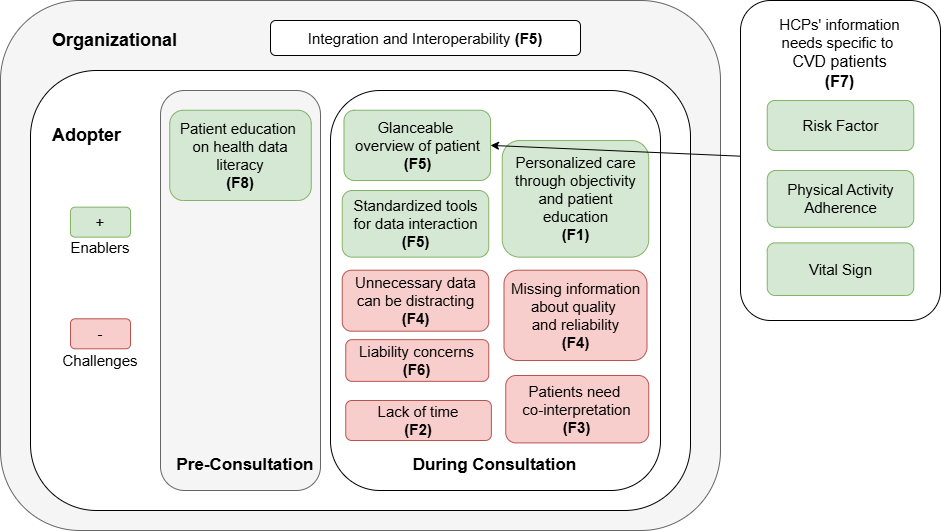}
    \caption{HCP perspectives on barriers and enablers of PGHD at different levels of the NASSS Framework.}
    \label{fig:design}
\end{figure}

We summarize the overall barriers and enablers faced by the HCPs in Figure \ref{fig:design}, arranged by different levels of the healthcare ecosystem as categorized by the NASSS framework for implementation science \cite{greenhalghAdoptionNewFramework2017} to better contextualize and scaffold our findings. Key findings are labeled (e.g., F0, F1, ...) to facilitate reference and linkage with the discussion section.

% \begin{figure}
%     \centering
% \includegraphics[width=0.9\linewidth]{images/DesignOpportunities.png}
%     \caption{HCP Perspectives on barriers and enablers of PGHD at different levels of the NASSS Framework}
%     \label{fig:design}
% \end{figure}

\subsection{Part 1 - Situated Study Outcomes}

\subsubsection{Demographics}
Part 1 included 6 participants (P; 1M, 5F; 36 - 69 years (M = 43, SD = 11.6, Median = 40 yrs) of age. 2 P have completed high school and 4 university-level education. 4 P reported being regular users of activity tracking devices or applications. Based on the RAPA questionnaire \cite{topolskiPeerReviewedRapid2006}, none of them were considered as sedentary. They engaged in light or moderate activities in a weekly basis.

\subsubsection{Integrating Objective Data to Personalize Care and Educate Participants - (F1)}

We observed that HCPs began consultations by engaging participants in discussions about their health knowledge, such as the importance of physical activity for cardiovascular health: \emph {``What do you know about the importance of physical activity in cardiovascular health and your heart health?''} (HCP1 with P6). To gain a comprehensive understanding of physical activity patterns and wellbeing, HCPs referred to both printed reports and data from the application, explaining, \emph {``We use the wearables and the sensors just to get better data about your activity pattern''} (HCP with P6). By combining objective data from wearables with subjective input about preferences and habits, HCPs developed detailed profiles of participants’ risks, behaviours, and barriers to exercise. Based on these insights, they collaboratively engaged in planning for the integration of preferred physical activities into participants’ lifestyles.

However, HCPs highlighted the challenges with relying on subjective data, particularly when assessing training intensities or dealing with patient forgetfulness. HCP1 highlighted how patients often fail to recall their physical activity history accurately, while HCP2 emphasized the limitations of subjective reporting in physical activity planning consultations. The integration of objective PGHD could address these challenges, offering a more trustworthy alternative to patient-reported activity levels. Finally, HCP2 highlighted the impact of PGHD, emphasizing that it allows patients to see their own unique data reflected in consultations, fostering a more personalized and meaningful consultation between patients and HCPs.

%Participants also recognized the added value and authenticity of objective data, with P1 and P3 stressing how it provides a factual basis—P3 noted, \emph {``you cannot say, I do 10,000 steps every day when you don’t do it.''} This objectivity is essential, as P1 pointed out that self-reported information are often inaccurate. 

\subsubsection{HCPs' need for time-efficient engagement with data (F2)}

Time constraints were a significant barrier for HCPs engaging PGHD in detail. While HCP1 acknowledged that PGHD could enhance patient-centred care, they also noted that it introduces complexity and extends consultation durations, creating challenges in the context of tightly scheduled appointments. They highlighted that wearable data, for instance, often triggers additional discussions, adding layers of complexity to the limited time available for each patient:  \emph {``It brings another layer of complexity ... you have certain time slots, and the next patient is waiting.''} Moreover, navigating mobile applications to extract relevant data posed further difficulties, as HCP1 explained, \emph {``[a]s a physician…because I didn’t know the applications it was hard to ... draw the data that is relevant for me.''} Similarly, HCP2 emphasized that many colleagues often dismiss PGHD altogether due to time constraints due to additional time required for recontextualizing PGHD, remarking, \emph {``They say, ‘Well, I don’t have time to look at this,’ so it’s not accepted.''} 

Furthermore, activities recorded by participants often lacked sufficient detail or contain inaccuracies, requiring HCPs to prompt participants for clarification. For instance, P2 needed several minutes to recall specifics about their most active day, while P6 clarified that football playing had been recorded as running. Beyond objective data, HCPs also gathered subjective information, including personal preferences, daily routines, and barriers to physical activity, to provide meaningful recommendations. This multifaceted engagement highlights the need for time-efficient approaches to effectively use PGHD for getting a holistic picture of patients.

\subsubsection{HCPs' role as co-interpreters of PGHD (F3)}

The participants often relied on HCPs to interpret their data, highlighting their role in reducing anxiety and confusion over irregular readings. For example, P1 described feeling panicked by fluctuating device measurements, emphasizing the need for a doctor to explain these variations: \emph {``I think you need a doctor to talk you through the numbers as well ... I panic a little bit for a good day I was like am I gonna die like do I have asthma and then the next day it was like oh your oxygen [level] is fine.''} P4 underscored the importance of personalized advice from HCPs, particularly regarding physical activity, saying, \emph {``I feel good and [it] helped me when he explained me about what kind of exercises to do, keep in mind my heart, [and] not to [exceed my heart rate] for a certain exercise.''}  This sentiment aligns with feedback from HCPs during the workshop, where they highlighted clarifying with patients about discussing abnormal readings and addressing common patient misunderstandings, such as confusing heart rate and blood pressure (c.f. Section \ref{sec:bpHR}). These examples illustrate how participants and HCPs collaboratively interpret PGHD to reduce confusion, clarify irregular readings, and provide tailored guidance for consultations.

\subsubsection{HCPs Need Concise, Reliable, and Relevant Data (F4)}
HCPs aimed to develop a holistic understanding of their patients while avoiding data overload during consultations. They recognize that PGHD can offer valuable insights, particularly regarding fitness and physical activity levels. As HCP2 remarked, \emph {``PGHD will [give a] relatively good picture of the fitness and physical activity of a patient.''} However, data needs vary significantly among HCPs, with a clear preference for health and physical activity history \cite{serbanJustSeeNumbers2023}. Irrelevant or excessive data is often viewed as a barrier, as illustrated by HCP1, who emphasized the importance of focusing on key variables such as physical activity, weight, and blood pressure, stating, \emph {``Everything else...is kind of a distraction.''} Concerns about the reliability of data generated by wearables also arose, with distrust expressed toward certain biomarkers, such as ``body composition`'', which HCP1 described as unreliable.

HCPs also highlighted the importance of transparency in understanding when patients are actively recording their data. HCP2 proposed a ``data quality panel'' within applications to help professionals assess reliability and completeness, suggesting that such a panel could display metrics like data coverage over a specific period. HCP2 explained, \emph {``I would build a data quality panel on top, where you can see 95\% of data were captured... such a data quality panel...is super important.''} Less tech-savvy physicians also face challenges in quickly assessing data reliability, as noted by HCP1. These insights highlight the need for digital tools that provide clear, concise summaries of data quality, enabling HCPs to determine whether they have a complete and reliable picture of their patients' health for effective SDM.

\subsubsection{Need for Standardized PGHD Interfaces (F5)}

HCPs emphasized the importance of standardized interfaces and technologies to streamline data sense-making and analysis during consultations. HCP2 highlighted the usefulness of printed reports, stating that they were easier to navigate with patients compared to the smartphone interfaces, which may cause usability issues due to inconsistencies across devices: \emph {``it always happens that the next moment you have the weather report because you did something wrong there.''}. This highlights the challenges of managing diverse device interfaces in clinical practice. Furthermore, our observations revealed that HCPs commonly relied on the summarized printed reports for an overview while using mobile applications for deeper insights into the participants activity history and vital signs. HCP2 also stressed the need for a standardized visualization interface, suggesting that PGHD should be displayed consistently, regardless of the tracking device used.

Integrated data visualization emerged as a key enabler for identifyingtrends in self-tracked data and facilitate more personalized consultations. HCP1 expressed that visualizing the relationship between physical activity and weight loss could help uncover connections between biomarkers and support personalized consultation with patients. This would in turn allow them to clearly demonstrate how increased activity improves cardiovascular health. 

Finally, HCP1 proposed improved collaboration between medical device departments and physicians, suggesting a workflow where the device department preselects appropriate sensors and variables for patients.

\subsubsection{HCPs' liability concerns (F6)}

Finally, HCPs also expressed concerns about the clinical validity of PGHD and its implications for medical liability. When PGHD originates from non-certified medical devices, HCPs had to exercise caution in using this data for clinical decision-making. HCP1 highlighted: \emph {``There is a bit of liability because these are not medical devices, so I cannot change, for example, a blood pressure medication based on the data...There’s a gray area of liability... can I change my medical decision-making based on these data or not? So that’s a huge problem these days.''} These concerns highlight the need for a balance between accessing reliable and actionable PGHD and reducing the risk of exposing HCPs to malpractice risks.

\subsection{Part 2 - Card Sorting Workshop Outcomes}
\subsubsection{Broad information needs and interests of HCPs}

We identified three major classes of objective information that the HCPs wanted for effective shared decision making and physical activity goal setting \textbf{(F7)}:
\begin{enumerate}
    \item Monitoring vital signs - Measurements such as blood pressure and heart rate
    \item Adherence to physical activity -  Evaluating user's physical activity levels using level of moderate to vigorous physical activity (MVPA) levels, step counts, sedentary time, physical activity type and duration. 
    \item Risk factor monitoring - Assessing measures such as body mass index (BMI), sleep duration and quality, weight. 
\end{enumerate}
Building upon these, the HCPs wanted to gather subjective information from participants such as enablers and barriers to physical activity before crafting the physical activity plan. Furthermore, we observed that they wanted to have access to multi-modal data, such as blood pressure measurement, heart rate, physical activity, step count and sleep and had diverse data representation and analytics needs based on the three scenarios. We summarise the specific information needs from the perspectives of HCPs below. Details on the output of the card sorting activity are in Appendix \ref{appendix:scenario1}.

\subsubsection{Scenario 1: How can patients reflect on their PGHD before physical activity planning sessions?}

HCPs emphasized the need for clear definitions and explanations of vital signs and data points to enable patient reflection before physical activity planning sessions \textbf{(F8)}. \label{sec:bpHR} HCP1(WS) noted that patients unfamiliar with health tracking need \emph{``an optimal range and...cut-offs [about what] is normal, [what] is not normal, ...''} along with explanations of terms such as blood pressure and heart rate, as patients often confuse them.

\subsubsection{Scenario 2: How can HCPs have an overview on a patient before a physical activity planning session? }

The HCPs expressed a need for a quick and holistic overview of patients before physical activity planning sessions. This includes patient summaries \cite{ElectronicCrossborderHealth2024}, as well as key information extracted from electronic health records (EHR). Specific details, such as maximal heart rate (MaxHR) — which requires clinical procedures for fully valid observation — were identified as critical for decision-making. HCPs also highlighted the importance of understanding the patient’s medical history, current goals, and barriers or enablers, while preferring to address some of these aspects, such as barriers, during the consultation itself. HCP2(WS) emphasized the need for essential metrics like resting heart rate, MaxHR, sedentary time, and physical activity duration. Finally, they also stressed the importance of aligning planning not only with general guidelines but also with individualized goal achievement. These findings underscore the need for tailored tools that integrate clinical EHR data and PGHD to support decision making during physical activity planning. These responses do not reflect isolated expert opinions, but build directly on their reflections from Part 1, where HCPs collaborated with participants using PGHD in situ.

\subsubsection{Scenario 3: How can PGHD support shared decision making during physical activity planning sessions?}

Both HCPs identified Scenario 3 as the most crucial among the scenarios presented during the workshop, emphasizing its role in achieving actionable outcomes through PGHD supported SDM. They highlighted the value of collaboratively reviewing data with patients during consultations, provided that the visualizations are clear and enable quick analysis. This approach allows HCPs to inform patients, discuss findings, collect feedback, and set new goals together, fostering a patient-centred planning process. HCP1(WS) emphasized the significance of reviewing data during consultations rather than beforehand, stating that the key is to have the data accessible for joint discussions and goal setting.

While the information requirements remained consistent across consultations, HCP1(WS) noted the necessity of knowing MaxHR during the first session to tailor the discussion effectively. \textit{We recognize that in real-world cardiac rehabilitation, patients often undergo center-based sessions where parameters like MaxHR are already addressed. However, our aim was to explore potential scenarios of PGHD shared by patients across both initial and ongoing consultations. Hence, this scenario was designed to surface HCP information preferences in diverse consultation settings, including first encounters or hybrid models, which could help with decision-making} 

In terms of trustworthiness, HCP1(WS) regarded brand information as potentially helpful but not essential, provided the devices functioned reliably and patients could manage them effectively.

The HCPs also proposed a structured flow for data-enabled SDM, starting with discussing physical activity metrics such as MVPA, steps, and sedentary time, followed by correlations with health indicators like blood pressure, BMI, weight, and sleep. The discussion would then shift to identifying enablers and barriers to physical activity and creating actionable plans. HCP1(WS) and HCP2(WS) emphasized the importance of promoting health literacy and intrinsic motivation to help patients achieve their goals, while another suggested focusing on barriers and enablers to guide planning. Additionally, they suggested introducing a table visualization to document patient-specific barriers and enablers, which could serve as a tangible artifact to support SDM during planning sessions. Although the proposed steps align with the flows observed during contextual inquiries, the HCPs acknowledged that actual consultations with patients might be of greater fluidity due to their dynamic nature.

\subsubsection{Key information for HCPs}

Both HCPs agreed that unnecessary data often distracted them during consultations, as highlighted in study Part 1. Through the MoSCoW prioritization activity (Figure \ref{fig:moscow}), HCPs identified the types of data essential for the physical activity planning process. Specifically, they emphasized the need for information about the devices used by patients to collect PGHD, such as the brand, model, and medical equipment certification. The reliability of data was a significant concern, particularly for certain metrics like blood pressure. \textit{Although participants in Part 1 used Withings devices exclusively, this context was not emphasized to workshop participants in Part 2. HCPs responded based on general clinical experience, where patients may bring data from various sources with unknown certification. This lack of priming was to surface real-world challenges in PGHD interpretation.}

As HCP1(WS) noted, some wearable devices are often inaccurate for blood pressure measurements, making it critical to understand the device used. However, this level of detail was not required for all metrics; for example, information about the brand of a weight scale was deemed unnecessary. HCP2(WS) further highlighted the importance of knowing details like sleeve width for blood pressure devices. These findings underscore the need to present device-specific characteristics required by HCPs for assessing the reliability of the data.

\subsubsection{Overall Enablers and Barriers}

HCPs identified multi-level factors affecting PGHD integration into clinical workflows, similar to findings from Part 1. Among benefits, HCPs noted that tracking PGHD can provide CVD patients with a sense of safety, as it allows them to monitor key metrics such as heart rate and blood pressure. As HCP2(WS) remarked, ``The patients feel safer when [they] can track it and can see okay my heart rate is in range, my blood pressure is in range.'' This aligns with existing literature showing that CVD patients may experience kinesophobia — the fear of physical activity — following an acute cardiac event, which can lead to reduced activity and non-adherence \cite{keessenFactorsRelatedFear2020}.
However, HCPs also identified several barriers across organizational, legal, and workload domains. At the organizational level, HCPs emphasized the lack of systems to effectively integrate PGHD into their workflows. As HCP1(WS) explained, “We do not have systems to integrate such data... patients come with their own wearables, and we say, yeah, that’s great, but we cannot use it really and integrate it really.” Furthermore, they expressed their perceptions of how data privacy regulations at the state and national levels further complicated integration. HCP1(WS) highlighted, \emph{``We have the legal side, but it’s in the data privacy... the framework of the state, Austria, and Salzburg.''}
Workload considerations also play a role, as educating patients on what to track might require significant time during consultations. HCP2(WS) noted, \emph{``It’s hard to integrate it into everyday life... you have to explain to them what they have to track.''} Additionally, personalizing healthcare for patients with multiple conditions presents unique challenges, as each patient may require tailored metrics. HCP1(WS) observed, \emph{``For cardiovascular or heart disease patients it may be different... Maybe I have to focus on weight for one patient and on heart rate and physical activity for another.''}

Overall, both HCPs agreed that the integration of PGHD into clinical workflows for physical activity planning faces multi-level challenges that require addressing organizational challenges, data privacy concerns, and patient education to achieve effective implementation. While concerns around data privacy and security are well-documented,\textit{ our aim was to highlight how such concerns manifest at the point of care from the HCP’s perspective, which remains underrepresented in design-oriented PGHD literature especially in the context of physical activity planning in cardiac rehabilitation. We acknowledge that further triangulation with legal and privacy experts and end users - patients - is necessary and propose this as a direction for future work.}
\section{Discussion}

Digital tools can support HCP-patient collaboration in chronic disease management \cite{chungMoreTelemonitoringHealth2015}. Prior work has explored integrating shared decision-making (SDM) in digital tools to engage patients in cardiovascular rehabilitation beyond planning sessions \cite{wurhoferInvestigatingSharedDecisionmaking2024,bonneuxSharedHeartApproachTechnologySupported2022,hoppchenTargetingBehavioralFactors2024}. However, HCI research on using PGHD for supporting patient-provider collaboration in physical activity planning - a key component of cardiac rehabilitation - is limited. The introduction of PGHD can be a double-edged sword — objectifying data while shifting roles, responsibilities, and consultation dynamics. \cite{jongsma_how_2021}. In the case of cardiac rehabilitation, HCPs may see potential in PGHD, but express concerns about data reliability and accuracy and the additional burden it might cause on their already time-constrained workflows \cite{alaboudCliniciansPerspectivesUsing2022}. To meaningfully support clinicians \cite{scholich_augmenting_2024} and patients with shared decision making, it is necessary to understand data based tasks that HCPs perform during physical activity planning consultations for cardiac rehabilitation, and design tools and processes that are \textit{fit for purpose}. This requires a deeper, contextual understanding of clinical practices, organizational constraints and interpersonal dynamics between HCPs and Patients.

To scaffold our interpretation, we draw on the NASSS framework \cite{greenhalghAdoptionNewFramework2017} which emphasizes that the success or failure of digital health innovation depends on multiple interacting domains - the technology, the adopters, the organization, and the broader system context. Although our study is set in a pre-clinical context with HCPs and healthy subjects, our formative design exploration provides more nuanced understandings on the challenges, enablers, and design considerations that are highly relevant for cardiac rehabilitation settings. The following subsections are organized around key insights (F1, F2, etc.) to link back to our findings.

\subsection{Supporting the Operationalization of PGHD for HCPs}
\subsubsection{Addressing Data Quality, Reliability and Trust Issues - (RQ1 and RQ1.2) \textbf{(F4)}}

PGHD can enhance SDM by providing additional insights about patients to HCPs when combined with information from EHRs. However its quality  significantly impacts HCPs’ confidence in conducting assessments \cite{alaboudCliniciansPerspectivesUsing2022} and integration into clinical workflows \cite{west2018common}. HCPs viewed data quality through multiple lenses, particularly focusing on the reliability and completeness of data to assess trustworthiness. They valued PGHD’s objectivity but stressed the need for reliable devices, and clear regulatory status (e.g., CE marking, MDR, FDA) to avoid the legal “grey zone” that limits use in diagnosis or treatment and increases liability \cite{khatiwada_patient-generated_2024,mcgraw2013going}. HCPs also emphasized the importance of data completeness, questioning how much data was recorded versus missing.

\emph{\textbf{Design Implication (RQ 2)}: 
\begin{itemize}
    \item Data quality information should include 1) Source of PGHD, 2) Medical device certification information (e.g. CE marked for which intended use) and 3) Completeness of data. This information must be presented in a quick and accessible format to enhance the trustworthiness of PGHD and allow HCPs to assess if they can incorporate the data for decision-making.
\end{itemize}
}

\subsubsection{HCP Information Needs for PA planning (RQ1.2) \textbf{(F7)}}

As ``information equals data plus meaning'' \cite{rumboldWhatAreData2018a}, HCPs engaged in a process of collective meaning-making by making sense of PGHD through dialogue and contextualization \cite{rutjesBenefitsCostsPatient2017}. To facilitate PA planning, HCPs identified four key information needs: vital signs, risk factors, physical activity adherence and habits, and enablers and barriers to PA \cite{hoppchenTargetingBehavioralFactors2024}. While the first three categories are primarily objective, the fourth includes subjective elements. It was important to ensure these information were conveyed quickly and effectively to clinicians, as [unnecessary] data often proved distracting during consultations, creating extra data work \cite{mollerWhoDoesWork2020}. Furthermore, the presence of comorbidities or varying medical conditions necessitated adaptable information tailored to each patient's risk factors and health status. This arguably underscores the need to expand upon Chung \emph{et al.}'s \cite{chung2016boundary} concept of goal-oriented visualization, moving toward patient-oriented visualizations and adaptable data sense-making capabilities to support personalized healthcare delivery \cite{chungMoreTelemonitoringHealth2015,diogo_2024}.

%SCR B6.3

\emph{\textbf{Design Implications (RQ 2)}: 
\begin{itemize}
    \item Visualizations should prioritize HCP's primary information needs: 1) Vital signs 2) Risk factors, 3) Physical activity adherence and habits, 4) Enablers and barriers to physical activity. 
\item Dashboards should be adaptable \cite{diogo_2024}, using templates for similar patient profiles
\end{itemize}}

\subsubsection{Supporting HCPs' Deeper Investigation Needs (RQ1.2)\textbf{ (F5)}}

HCPs expressed the need for an overview of the PGHD and the flexibility to explore data in depth. This would enable them to analyze information and demonstrate potential correlations between biomarkers to patients, thereby personalizing consultations \cite{alaboudCliniciansPerspectivesUsing2022}. The printed report from the Withings application provided an overview, and the mobile application partially supported this workflow., but usability was hindered by unfamiliarity with the interface, operating system, and small screen size. These challenges reflect a broader issue with consumer devices — limited integration with clinical systems and interface customization to fit workflows and HCP preferences.

\emph{\textbf{Design Implications (RQ 2)}: 
\begin{itemize}
\item Offer glance-able dashboards \cite{tadas_user-centred_2022} with pre-defined analysis functions and visualizations tailored to physical activity planning \cite{chungMoreTelemonitoringHealth2015}, while allowing deeper exploration of data when needed.
\item Offer ability to generate printed reports \cite{rajUnderstandingIndividualCollaborative2017b} offering HCPs more choice in interface preferences.
\end{itemize}
}

\subsubsection{Towards Artificial Intelligence (AI) Augmented Data work and Sense-making (RQ1) \textbf{(F1 and F2)}}

Our study emphasizes the uniqueness of individual patient needs, as well as the dynamic nature of HCP and patient-HCP interactions. This dynamism in workflows and requirements poses significant challenges for designing ``one-size-fits-all'' tools that integrate PGHD into physical activity planning workflows. However, AI tools offer promising potential to support sense-making in these evolving contexts \cite{davenportPotentialArtificialIntelligence2019a}. By generating ``goal-oriented or patient-oriented visualizations'' \cite{chung2016boundary}, AI may help in personalizing consultations, enabling HCPs to tailor their approaches to the specific context of each patient. Recent HCI research has further explored AI's role in facilitating SDM \cite{kimHowMuchDecision2024,haoAdvancingPatientCenteredShared2024,suWhatYourEnvisioned2022}, as well as in enabling individuals to reflect on their physical activity data through natural language interaction \cite{stromelNarratingFitnessLeveraging2024}.

\emph{\textbf{Design Implications (RQ 2)}: 
    \begin{itemize}
    \item Flexibility: Tools should provide HCPs with flexibility to select sense-making strategies that adapt workflows to patient or HCP-specific needs \cite{diogo_2024}.
    \item AI augmentation: AI can potentially augment data sense-making by reducing barriers to data analysis through features like summarization, trend identification, anomaly detection, and natural language interaction \cite{mastrianni2024ai}. These capabilities could significantly reduce resources required for data analysis, enabling HCPs to focus more effectively on patient-centred care.
    \end{itemize}
}

\subsection{Enabling Connected Health Systems}
%5.2.7,5.2.8
\subsubsection{Towards Seamless Data Integration (RQ1) \textbf{(F5)}}

HCPs perceived that patients have privacy concerns regarding sharing their personal health data. However, participants, though not patients, in our study were comfortable in sharing such data with HCPs to improve consultations. This is therefore pending validation with patients, but implies a need to align HCPs' and patients' perspectives on what data to track and share, as well as the necessity for connected health systems to integrate data from consumer wearables effectively \cite{oh_patients_2022,chung2016boundary,jayathissa_patient-generated_2023} and securely. HCPs also acknowledged that achieving such integration is challenging, requiring coordination at an national (e.g., GDPR) and organizational level on approved devices and working with health informatics teams to ensure seamless data integration into their workflows.

Addressing these challenges would offer two key benefits: First, HCPs' access to PGHD could be enhanced, enabling better preparation for consultations \cite{eunkyung_jo_geniauti_2022}. Second, presenting PGHD in a standardized format could reduce the time, effort, and cognitive load required for clinicians by minimizing the need to navigate multiple interfaces from various self-tracking applications.

\emph{\textbf{Design Implications (RQ 2)}: 
    \begin{itemize}
    \item Facilitate shared understanding: Systems must align HCPs and patients perspectives clarifying the purpose and value of PGHD. 
    \item Patients should be educated on best practices for self-tracking, interpreting data, and identifying data most relevant for physical activity planning. Digital patient companions \cite{hoppchenBeMeStay2024}, can streamline data integration, enhance data presentation for HCPs, filter out irrelevant information, and enable efficient, goal-oriented consultations \cite{chung2016boundary}.
    \item Standardize devices, data and privacy-compliant data pipelines \cite{giacomini_roadmap_2023}: Healthcare organizations and regulatory offices should establish an agreed-upon list of approved or recommended devices. This ensures interoperability, facilitates data and interface standardization, and reduces burdens for HCPs.
    \end{itemize}
}

\subsubsection{Being Mindful of Additional Data Work in Clinical Workflows (RQ1) \textbf{(F3 and F8)}}

Moller \emph{et al.} \cite{mollerWhoDoesWork2020} emphasized the importance of legitimizing data-centric tasks in clinical workplaces to ensure smooth integration of PGHD. We identified an emerging challenge in identifying who supports patients in developing health data literacy, enabling them to track and share data effectively. Unless tools appropriately automate preselection and preparation, patients must be informed about the specific data needs of clinicians during physical activity planning sessions before consultations, ensuring that the data they provide is relevant and actionable \cite{chung2016boundary,oh_patients_2022}. Adequate health data literacy is critical to help patients understand the vital signs they track, avoiding miscommunication with HCPs and improving self-management. Observations from our study reveal that HCPs frequently interpreted PGHD and guided patients on what and how to track, which can add significant burden to their existing workload \cite{haase2023data}.

HCP could be considerably more productive and efficient if PGHD e.g. obtained through a brief observation period before a first consultation can provide rich, reliable and concisely to-the-point information about their patients as they e.g. first see them. However, certain investments would arguably be required to this end. These efforts could be implemented at various levels, including organizational, healthcare systems, or service providers depending on the context to reduce burdens on HCPs during consultations.

\emph{\textbf{Design Implications (RQ 2)}: 
    \begin{itemize}
    \item Clearly identify who will educate patients on self-tracking. This could involve dedicated staff (e.g., nurses) or a digital companion \cite{hoppchenBeMeStay2024}, alleviating burden on HCPs during consultations. 
    \item As HCPs potentially accumulate additional data work as a result of PGHD integration \cite{mollerWhoDoesWork2020}, appropriate resources such as data analysis tools to augment their roles as data workers should be provided
    \end{itemize}
}

\subsection{The Need for Nuanced and Deeper Understanding of the Context (RQ1)}

Healthcare systems are recognized as complex adaptive socio-technical systems that require nuanced approaches to their design \cite{sittigNewSociotechnicalModel2010}. Andersen \emph{et al.} \cite{andersenAligningConcernsTelecare2019} emphasize that such systems must be designed with a deep understanding of the organizational context while addressing the needs of patients, HCPs, and other stakeholders \cite{waddellLeveragingImplementationScience2024a}. Effective digital health technologies demand that HCI researchers consider multiple system levels - micro, meso, and macro - as well as the stakeholders’ values \cite{borningNextStepsValue2012} and the artifacts facilitating information flows \cite{blandford2005dicot}, which are integral to understanding the healthcare context. 
While utilizing the NASSS framework \cite{greenhalghAdoptionNewFramework2017} helped us identify multilevel barriers to PGHD integration, our work emphasizes that empathy must extend beyond the interface and transcent into systemic workflows and socio-technical interactions.
With growing interest within the HCI community in bridging implementation science with HCI to enhance the design and deployment of digital health solutions \cite{lyonBridgingHCIImplementation2023,waddellLeveragingImplementationScience2024a}, our work paves the path for such interdisciplinary exchange as an enabler for effectively implementing digital health technologies in healthcare systems. 

\section{Limitations and Future Work}
Despite high ecological validity our formative study was conducted in a pre-clinical context with healthy participants based on situated digital health technology use and consultations. With 10 participants in total, the  generalizability of our findings are limited. Additionally, our analyses focus on HCP perspectives. However, we uncover preliminary challenges and enablers to integrating PGHD in clinical pathways for physical activity planning. As we focussed on the first physical activity planning session and participants only attended one session, we are not accompanying actual CVD patient pathways and where they tend to be generally older and have multiple risk factors or comorbidities. Furthermore, we instructed our participants to complete specific tracking tasks from devices from a single manufacturer, however actual patients' tracking habits could vary based on their medical conditions, technology acceptance, digital literacy and lifestyle. Further research is required to expand on our findings through a broader, longitudinal study exploring when / where / how to best offer analytics capabilities to both HCPs and patients and understand their impact on patient-provider collaboration with data and patient-centred care. By surfacing HCP practices in co-interpreting PGHD, our study lays the groundwork for future research on \textbf{long-term PA planning with follow-up consultations}, and focussing on \textbf{patient perspectives on} how such practices may enhance patient \textbf{agency}, \textbf{self-efficacy}, and readiness to overcome \textbf{kinesophobia}.
\section{Conclusion}

CVD remains the leading cause of death globally, and physical activity planning is crucial for secondary prevention. PGHD offers HCPs an objective and holistic picture of CVD patients allowing for personalized care. However its integration within the context of physical activity planning for cardiovascular rehabilitation remains unexplored. We conducted a two-part formative design exploration with HCPs and healthy subjects to gain deeper understandings on the context and the problem. Our empirical findings reveal that data sense-making and actionability is key for SDM, however there are several multi-level challenges in integrating, and presenting, the right amount of trustworthy data for time-limited HCPs for effective goal-oriented sense-making. Reflecting on our findings and contextualizing with related works, we offer practical design implications for addressing the arising challenges. Our work underscores the importance of considering perspectives beyond the primary stakeholders and across clinical settings when designing PGHD integration. Future research will include CVD patient perspectives on PGHD integration in physical activity planning consultations to design seamless patient journeys and clinical workflows which allows for better patient-HCP collaboration with data. Theories such as the distributed cognition theory \cite{zhang2021Data,blandford2005dicot} can be adopted to better understand how patients and HCPs interact with PGHD within a broader socio-technical system.

\section*{Authors' Contributions}
PVSP was the lead author and investigator. IH co-analyzed Part 1, and HM co-facilitated workshops and co-analyzed Part 2 data. MS and GT facilitated clinical site access, coordinated with stakeholders, and advised on procedures. IH and JDS contributed to manuscript writing. DW, JN, AS contributed to manuscript revisions and editing. JDS supervised PVSP and contributed to research methodology and conceptualization.


\newpage
\appendix
\section{Appendix}
\label{appendix:scenario1}

\begin{figure}
    \centering
    \includegraphics[width=1\linewidth]{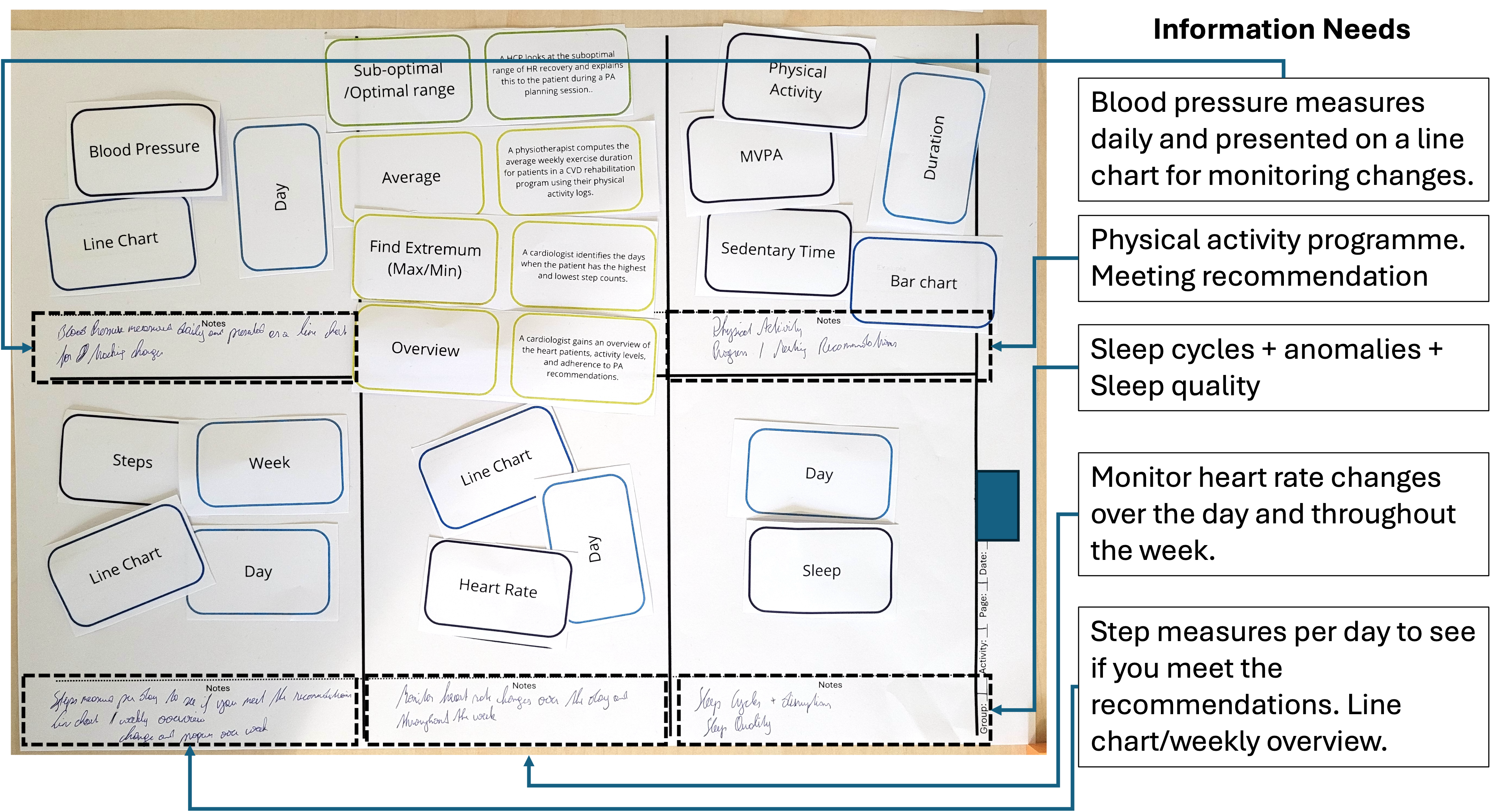}
    \caption{Scenario 1: Information that patients can reflect on before consultation}
    \label{fig:scenario1}
\end{figure}

\begin{longtable}{|p{3.5cm}|p{2.5cm}|p{2cm}|p{1.5cm}|p{3.5cm}|}
\hline
\multicolumn{5}{|l|}{\textbf{Scenario 1: How can patients reflect on their PGHD before PA planning sessions?}} \\
\hline
\textbf{Information Needs} & \textbf{Visualization} & \textbf{Data} & \textbf{Time} & \textbf{Analysis Cards} \\
\hline
Blood pressure measures daily and presented on a line chart for monitoring & Line Chart & Blood Pressure & Day & 1) Sub optimal/Optimal Range 2) Average 3) Find Extremum (Max/Min) 4) Overview \\
\hline
Physical activity programme. Meeting recommendation & Bar Chart & Physical activity, MVPA, Sedentary Activity & - & 1) Sub optimal/Optimal Range 2) Average 3) Find Extremum (Max/Min) 4) Overview \\
\hline
Sleep cycles + Anomalies + Sleep quality & - & Sleep Quality & Day & 1) Sub optimal/Optimal Range 2) Average 3) Find Extremum (Max/Min) 4) Overview \\
\hline
Step measures per day to see if you meet the recommendations. Line chart & Line Chart & Steps & Day & 1) Sub optimal/Optimal Range 2) Average 3) Find Extremum (Max/Min) 4) Overview \\
\hline
Monitor heart rate changes over the day and throughout the week & Line Chart & Heart Rate & Day & 1) Sub optimal/Optimal Range 2) Average 3) Find Extremum (Max/Min) 4) Overview \\
\hline

\end{longtable}

\newpage

\label{appendix:scenario2}

\begin{figure}
    \centering
    \includegraphics[width=1\linewidth]{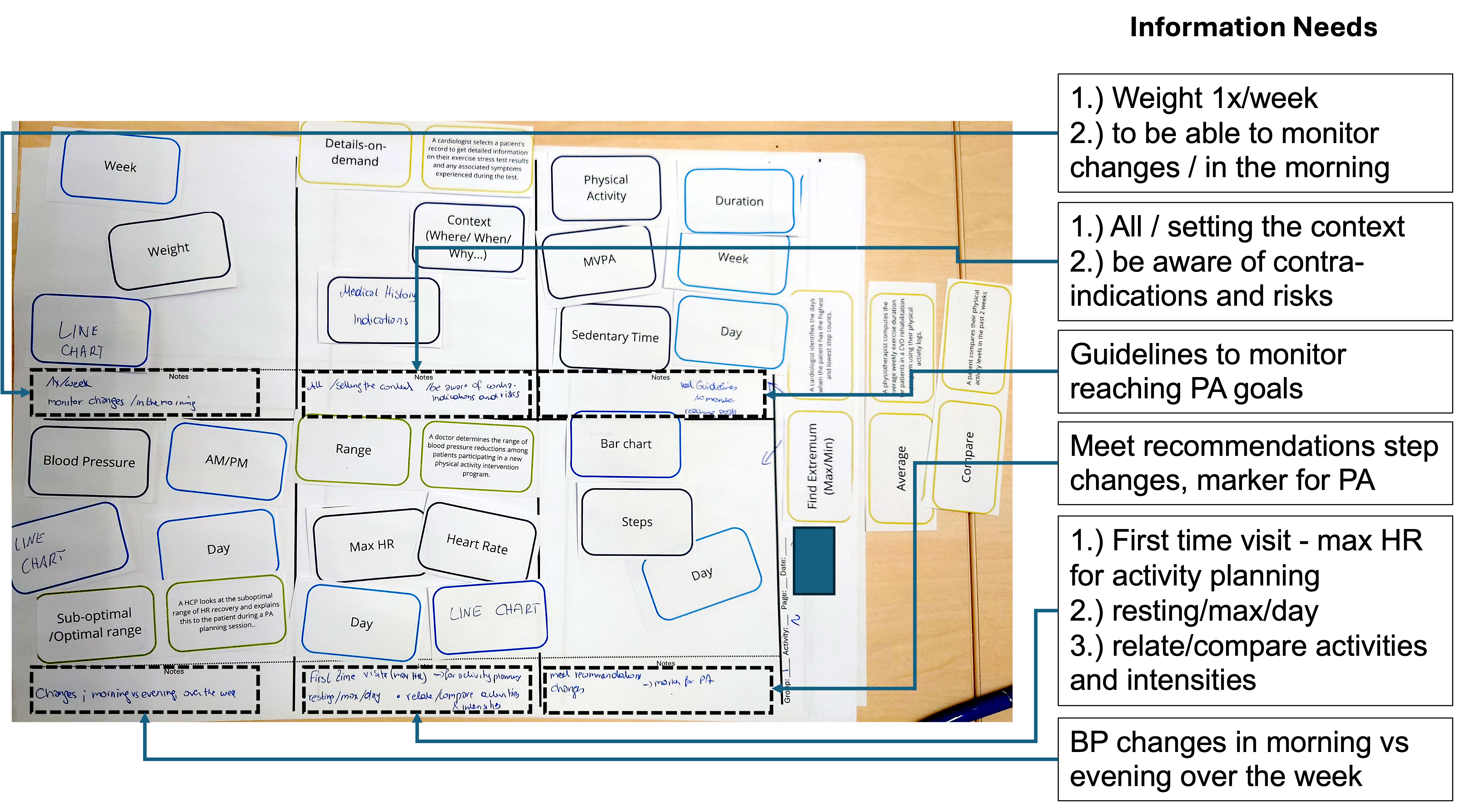}
    \caption{Scenario 2: HCPs information needs prior to consultations}
    \label{fig:scenario2}
\end{figure}
\begin{longtable}{|p{3.5cm}|p{2.5cm}|p{2cm}|p{1.5cm}|p{3.5cm}|}
\hline
\multicolumn{5}{|l|}{\textbf{Scenario 2: How can HCPs have an overview of the patients before PA planning sessions?}}\\
\hline
\textbf{Information Needs} & \textbf{Visualization} & \textbf{Data} & \textbf{Time} & \textbf{Analysis Cards} \\
\hline
Weight 1.) 1x/week 
2.) to be able to monitor changes / in the morning & Line Chart & Weight & Week &  \\
\hline
Changes in morning vs evening over the week & Line Chart & Blood Pressure & AM/PM & 1) sub-optimal/ optimal range \\
\hline
1.) All / setting the context 2.) be aware of contra-indications and risks & - & context (where, when, why) & - & 1) details-on-demand (e.g., test results and symptoms during test) \\
\hline
1.) First time visit - max HR for activity planning 
2.) resting/max/day 
3.) relate/compare activities and intensities & Line Chart & Heart Rate & Day & 1) Max HR 2) range \\
\hline
Guidelines to monitor reaching PA goals, & Bar Chart & Physical activity, MVPA, Sedentary Activity & Day & 1) find extremum 2) average 3) compare \\
\hline
meet recommendations changes, marker for PA & Bar Chart & Steps & Day & 1) find extremum 2) average 3) compare \\
\hline
\end{longtable}

\newpage
\label{appendix:scenario3}

\begin{longtable}{|p{3.5cm}|p{2.5cm}|p{2cm}|p{1.5cm}|p{3.5cm}|}
\hline
\multicolumn{5}{|l|}{\textbf{Scenario 3: How can PGHD support SDM in PA planning?}}\\
\hline
\textbf{Information Needs} & \textbf{Visualization} & \textbf{Data} & \textbf{Time} & \textbf{Analysis Cards} \\
\hline

Workflow starting here. Analysing PA levels with patient. Reached goals?:  Making correlations to health statistics promoting health literacy (Step 1) & Bar Chart & Physical activity, MVPA, Sedentary Activity, Steps, Medical Equipment certification, Brand, Model & - & Correlate, History \\
\hline
Step 2 & Bar Chart & Blood Pressure & Day, AM/PM & Correlate, History, Sub-optimal/Optimal Range \\
\hline
Step 3 & Line Chart & Weight, BMI & Weekday & Correlate, History \\
\hline
Step 4 & Bar Chart & Sleep Quality & Day & Correlate, History \\
\hline
Setting new goals (Step 5) & Table & Enablers to PA, Barriers to PA & - & Find extremum (Max/Min) \\
\hline

\end{longtable}

\newpage

\label{appendix:moscowPrioritization}

\begin{figure}
    \centering
    \includegraphics[width=0.9\linewidth]{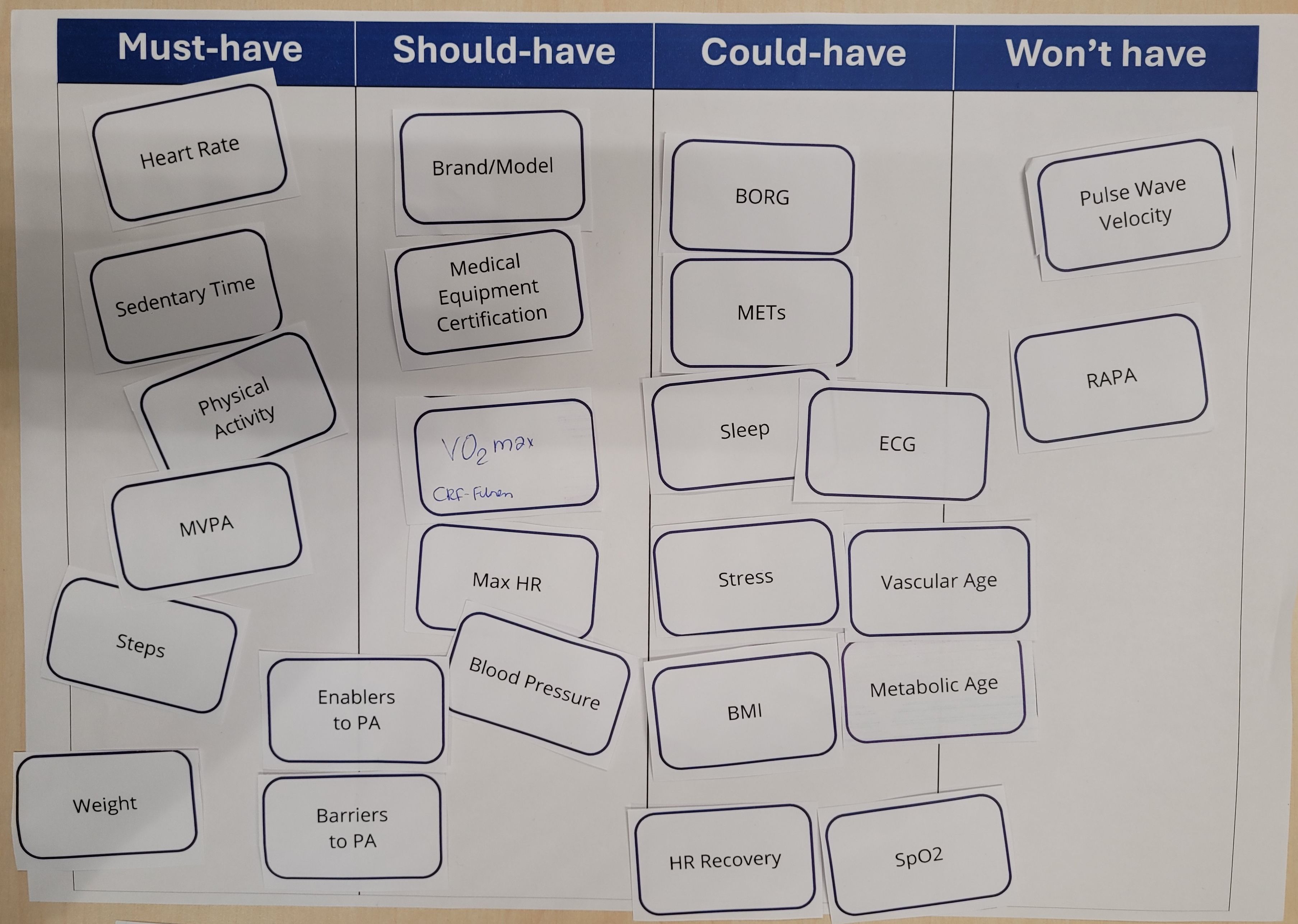}
    \caption{MoSCoW Prioritization of Data Points}
    \label{fig:moscow}
\end{figure}

\newpage

\label{appendix:analysis}
\begin{figure}
\centering
    \begin{minipage}{1\textwidth}
        \centering
\includegraphics[width=\linewidth]{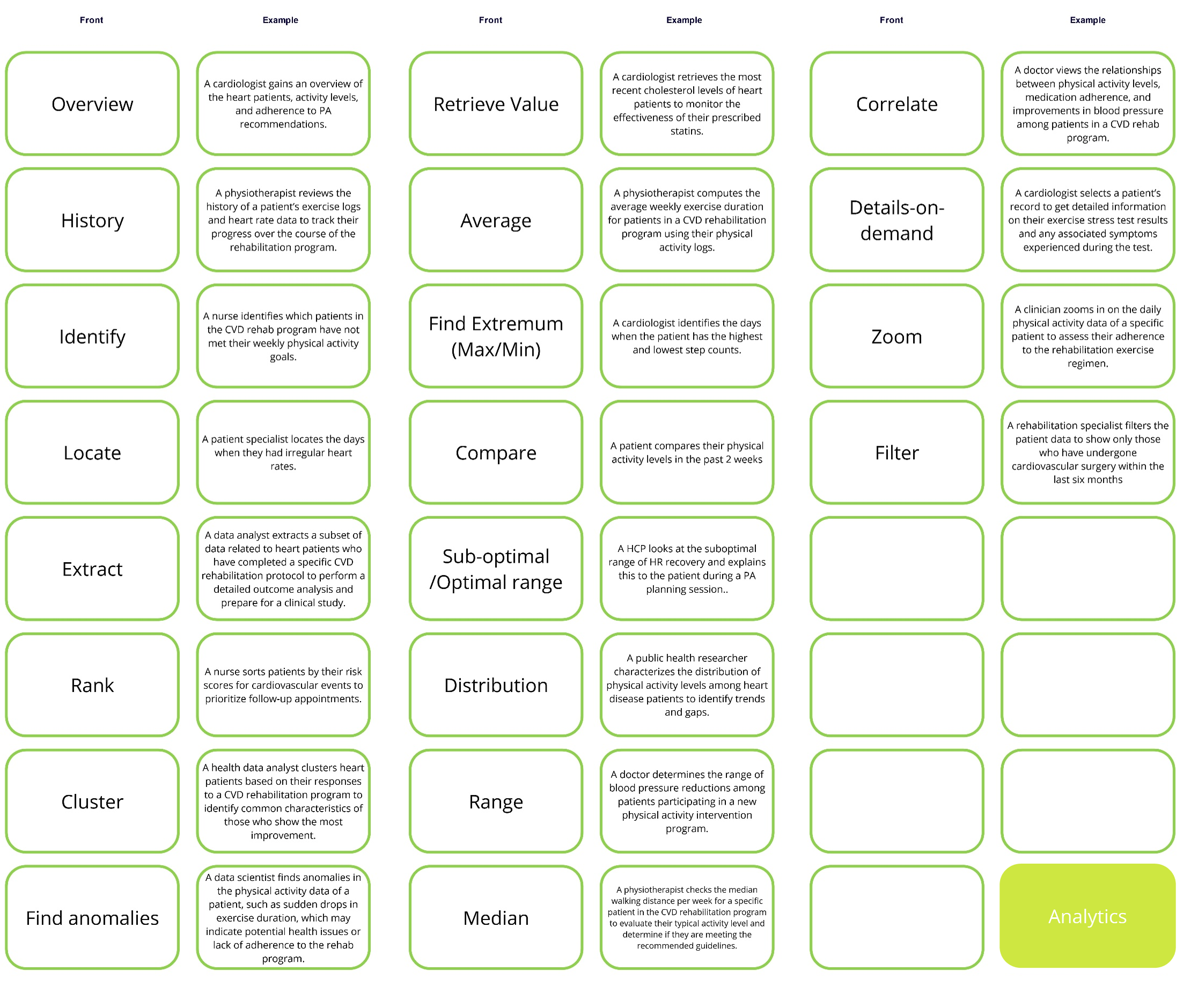}
    \end{minipage} 
\caption{Analysis Cards}
\label{fig:analysis_card}
\end{figure}

\label{appendix:data}
\begin{figure}
\centering
    \begin{minipage}{1\textwidth}
        \centering
\includegraphics[width=\linewidth]{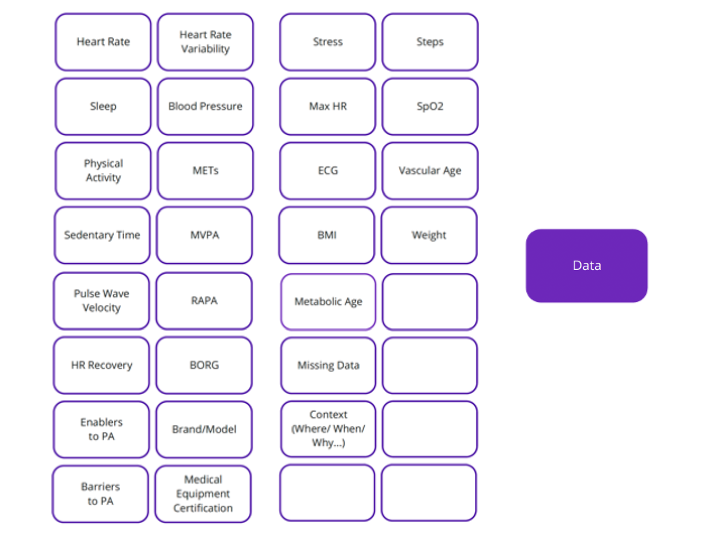}
    \end{minipage} 
\caption{Data Cards}
\label{fig:data_card}
\end{figure}

\label{appendix:time}
\begin{figure}
\centering
    \begin{minipage}{1\textwidth}
        \centering
\includegraphics[width=\linewidth]{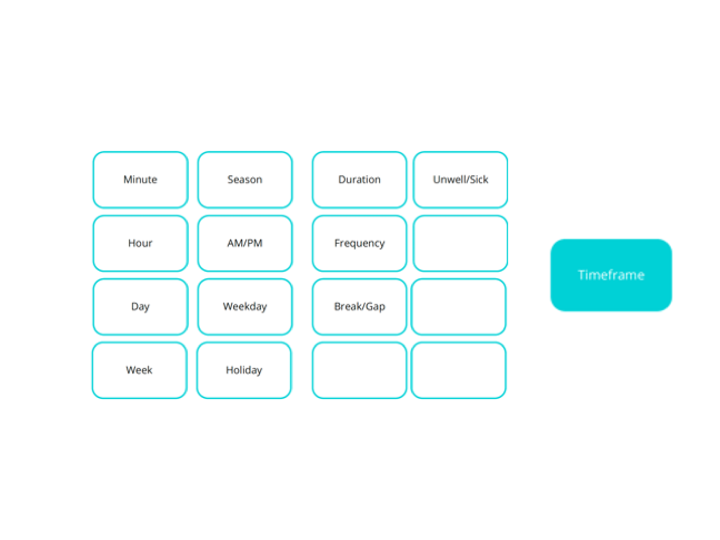}
    \end{minipage} 
\caption{Time Cards}
\label{fig:time_card}
\end{figure}

\label{appendix:visualization}
\begin{figure}
\centering
    \begin{minipage}{1\textwidth}
        \centering
\includegraphics[width=\linewidth]{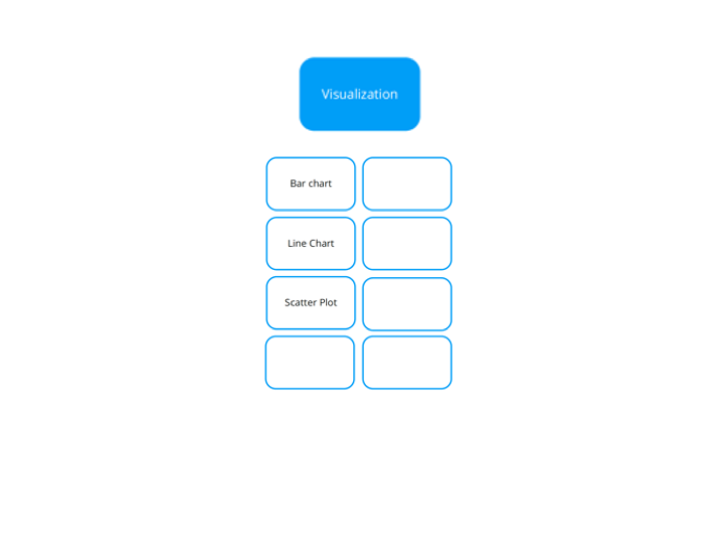}
    \end{minipage} 
\caption{Visualization Cards}
\label{fig:viz_card}
\end{figure}

\newpage
\
\bibliographystyle{splncs04}
\bibliography{0.sample-base}

\begin{thebibliography}{100}
\providecommand{\url}[1]{\texttt{#1}}
\providecommand{\urlprefix}{URL }
\providecommand{\doi}[1]{https://doi.org/#1}

\bibitem{alaboudCliniciansPerspectivesUsing2022}
Alaboud, K., Shahreen, M., Islam, H., Paul, T., Rana, M.K.Z., Morrison, A., Kumar, A., Mosa, A.S.M.: Clinicians’ perspectives in using patient-generated health data to improve ischemic heart disease management. AMIA Summits on Translational Science Proceedings  \textbf{2022}, ~112 (2022)

\bibitem{andersenAligningConcernsTelecare2019}
Andersen, T.O., Bansler, J.P., Kensing, F., Moll, J., Mønsted, T., Nielsen, K.D., Nielsen, O.W., Petersen, H.H., Svendsen, J.H.: Aligning concerns in telecare: Three concepts to guide the design of patient-centred e-health. Computer Supported Cooperative Work (CSCW)  \textbf{28}(6),  1039--1072 (Oct 2019)

\bibitem{aranki2016real}
Aranki, D., Kurillo, G., Yan, P., Liebovitz, D.M., Bajcsy, R.: Real-time tele-monitoring of patients with chronic heart-failure using a smartphone: lessons learned. IEEE Transactions on Affective Computing  \textbf{7}(3),  206--219 (2016)

\bibitem{ashur_wearable_2021}
Ashur, C., Cascino, T.M., Lewis, C., Townsend, W., Sen, A., Pekmezi, D., Richardson, C.R., Jackson, E.A.: Do {Wearable} {Activity} {Trackers} {Increase} {Physical} {Activity} {Among} {Cardiac} {Rehabilitation} {Participants}? {A} {SYSTEMATIC} {REVIEW} {AND} {META}-{ANALYSIS}. Journal of Cardiopulmonary Rehabilitation and Prevention  \textbf{41}(4), ~249 (Jul 2021)

\bibitem{DevelopPhysicalActivity}
Association, A.H.: Develop a {Physical} {Activity} {Plan} for {You}. \url{https://www.heart.org/en/health-topics/cardiac-rehab/getting-physically-active/develop-a-physical-activity-plan-for-you} (2024), accessed: 2024-05-31

\bibitem{Ayobi2017}
Ayobi, A., Marshall, P., Cox, A.L., Chen, Y.: Quantifying the body and caring for the mind: Self-tracking in multiple sclerosis. In: Proceedings of the 2017 CHI Conference on Human Factors in Computing Systems. CHI ’17, ACM (May 2017)

\bibitem{bauer2020implementation}
Bauer, M.S., Kirchner, J.: Implementation science: What is it and why should i care? Psychiatry research  \textbf{283},  112376 (2020)

\bibitem{bethellCardiacRehabilitationIt2008}
Bethell, H.J.N., Lewin, R.J.P., Dalal, H.M.: Cardiac rehabilitation: it works so why isn't it done? British Journal of General Practice  \textbf{58}(555),  677--679 (2008)

\bibitem{blandford2005dicot}
Blandford, A., Furniss, D.: Dicot: a methodology for applying distributed cognition to the design of teamworking systems. In: International workshop on design, specification, and verification of interactive systems. pp. 26--38. Springer (2005)

\bibitem{bonneuxSharedHeartApproachTechnologySupported2022}
Bonneux, C., Hansen, D., Dendale, P., Coninx, K.: The {SharedHeart} {Approach}: {Technology}-{Supported} {Shared} {Decision} {Making} to {Increase} {Physical} {Activity} in {Cardiac} {Patients}. In: Lewy, H., Barkan, R. (eds.) Pervasive {Computing} {Technologies} for {Healthcare}. pp. 469--488. Springer International Publishing, Cham (2022)

\bibitem{borningNextStepsValue2012}
Borning, A., Muller, M.: Next steps for value sensitive design. In: Proceedings of the {SIGCHI} {Conference} on {Human} {Factors} in {Computing} {Systems}. pp. 1125--1134. {CHI} '12, Association for Computing Machinery, New York, NY, USA (May 2012)

\bibitem{bossen2019data}
Bossen, C., Pine, K.H., Cabitza, F., Ellingsen, G., Piras, E.M.: Data work in healthcare: {An} {Introduction} (2019), number: 3 Pages: 465–474 Volume: 25

\bibitem{diogo_2024}
Branco, D., M\'{o}teiro, M., Bou\c{c}a-Machado, R., Miranda, R., Reis, T., Decoroso, E., Cardoso, R., Ramalho, J., Rato, F., Malheiro, J., Miranda, D., Cani\c{c}a, V., Pona-Ferreira, F., Guerreiro, D., Leit\~{a}o, M., Braz, A.S., J~Ferreira, J., Guerreiro, T.: Co-designing customizable clinical dashboards with multidisciplinary teams: Bridging the gap in chronic disease care. In: Proceedings of the 2024 CHI Conference on Human Factors in Computing Systems. CHI '24, Association for Computing Machinery, New York, NY, USA (2024)

\bibitem{braun2012thematic}
Braun, V., Clarke, V.: Thematic analysis. American Psychological Association (2012)

\bibitem{budd_burnout_2023}
Budd, J.: Burnout {Related} to {Electronic} {Health} {Record} {Use} in {Primary} {Care}. Journal of Primary Care \& Community Health  \textbf{14},  21501319231166921 (Apr 2023)

\bibitem{cernaChangingCategoricalWork2020}
Cerna, K., Grisot, M., Islind, A.S., Lindroth, T., Lundin, J., Steineck, G.: Changing {Categorical} {Work} in {Healthcare}: the {Use} of {Patient}-{Generated} {Health} {Data} in {Cancer} {Rehabilitation}. Comput. Supported Coop. Work  \textbf{29}(5),  563--586 (Oct 2020)

\bibitem{choeCharacterizingVisualizationInsights2015}
Choe, E.K., Lee, B., schraefel, m.: Characterizing {Visualization} {Insights} from {Quantified} {Selfers}' {Personal} {Data} {Presentations}. IEEE Computer Graphics and Applications  \textbf{35}(4),  28--37 (Jul 2015), conference Name: IEEE Computer Graphics and Applications

\bibitem{chungMoreTelemonitoringHealth2015}
Chung, C.F., Cook, J., Bales, E., Zia, J., Munson, S.A.: More {Than} {Telemonitoring}: {Health} {Provider} {Use} and {Nonuse} of {Life}-{Log} {Data} in {Irritable} {Bowel} {Syndrome} and {Weight} {Management}. Journal of Medical Internet Research  \textbf{17}(8) (Aug 2015), mAG ID: 1885665464

\bibitem{chung2016boundary}
Chung, C.F., Dew, K., Cole, A., Zia, J., Fogarty, J., Kientz, J.A., Munson, S.A.: Boundary negotiating artifacts in personal informatics: patient-provider collaboration with patient-generated data. In: Proceedings of the 19th ACM conference on computer-supported cooperative work \& social computing. pp. 770--786 (2016)

\bibitem{clegg1994case}
Clegg, D., Barker, R.: Case method fast-track: a RAD approach. Addison-Wesley Longman Publishing Co., Inc. (1994)

\bibitem{ElectronicCrossborderHealth2024}
Commission, E.: Electronic cross-border health services - {European} {Commission} (Apr 2024)

\bibitem{consolvoDesignRequirementsTechnologies2006}
Consolvo, S., Everitt, K., Smith, I., Landay, J.A.: Design requirements for technologies that encourage physical activity. In: Proceedings of the {SIGCHI} {Conference} on {Human} {Factors} in {Computing} {Systems}. pp. 457--466. {CHI} '06, Association for Computing Machinery, New York, NY, USA (Apr 2006)

\bibitem{costafigueiredoUsingDataApproach2021l}
Costa~Figueiredo, M., Su, H.I., Chen, Y.: Using data to approach the unknown: Patients' and healthcare providers? data practices in fertility challenges. Proceedings of the ACM on Human-Computer Interaction  \textbf{4}(CSCW3),  1--35 (2021)

\bibitem{craig_international_2017}
Craig, C., Marshall, A., Sjostrom, M., Bauman, A., Lee, P., Macfarlane, D., Lam, T., Stewart, S.: International physical activity questionnaire-short form. J Am Coll Health  \textbf{65}(7),  492--501 (2017)

\bibitem{craig_international_2003}
Craig, C.L., Marshall, A.L., Sjöström, M., Bauman, A.E., Booth, M.L., Ainsworth, B.E., Pratt, M., Ekelund, U., Yngve, A., Sallis, J.F., {others}: International physical activity questionnaire: 12-country reliability and validity. Medicine \& science in sports \& exercise  \textbf{35}(8),  1381--1395 (2003), publisher: LWW

\bibitem{damschroderFosteringImplementationHealth2009}
Damschroder, L.J., Aron, D.C., Keith, R.E., Kirsh, S.R., Alexander, J.A., Lowery, J.C.: Fostering implementation of health services research findings into practice: a consolidated framework for advancing implementation science. Implementation Science  \textbf{4}(1), ~50 (Aug 2009)

\bibitem{damschroder2022updated}
Damschroder, L.J., Reardon, C.M., Widerquist, M.A.O., Lowery, J.: The updated consolidated framework for implementation research based on user feedback. Implementation science  \textbf{17}(1), ~75 (2022)

\bibitem{davenportPotentialArtificialIntelligence2019a}
Davenport, T., Kalakota, R.: The potential for artificial intelligence in healthcare. Future Healthcare Journal  \textbf{6}(2),  94--98 (Jun 2019)

\bibitem{davis2014systematic}
Davis, M.M., Freeman, M., Kaye, J., Vuckovic, N., Buckley, D.I.: A systematic review of clinician and staff views on the acceptability of incorporating remote monitoring technology into primary care. Telemedicine and e-Health  \textbf{20}(5),  428--438 (2014)

\bibitem{ding2020effects}
Ding, H., Jayasena, R., Chen, S.H., Maiorana, A., Dowling, A., Layland, J., Good, N., Karunanithi, M., Edwards, I.: The effects of telemonitoring on patient compliance with self-management recommendations and outcomes of the innovative telemonitoring enhanced care program for chronic heart failure: randomized controlled trial. Journal of medical Internet research  \textbf{22}(7),  e17559 (2020)

\bibitem{duran2023applying}
Duran, A.T., Keener-DeNoia, A., Stavrolakes, K., Fraser, A., Blanco, L.V., Fleisch, E., Pieszchata, N., Cannone, D., McKay, C.K., Whittman, E., et~al.: Applying user-centered design and implementation science to the early-stage development of a telehealth-enhanced hybrid cardiac rehabilitation program: Quality improvement study. JMIR Formative Research  \textbf{7}(1),  e47264 (2023)

\bibitem{eunkyung_jo_geniauti_2022}
{Eunkyung Jo}, {Seora Park}, {Hyeonseok Bang}, {Youngeun Hong}, {Yeni Kim}, {Jungwon Choi}, {Bung Nyun Kim}, {Daniel A. Epstein}, {Hwajung Hong}: {GeniAuti}: {Toward} {Data}-{Driven} {Interventions} to {Challenging} {Behaviors} of {Autistic} {Children} through {Caregivers}' {Tracking}. Proceedings of the ACM on human-computer interaction  \textbf{6}(CSCW1),  1--27 (Mar 2022), mAG ID: 4226267185

\bibitem{fiske2019health}
Fiske, A., Buyx, A., Prainsack, B.: Health information counselors: a new profession for the age of big data. Academic Medicine  \textbf{94}(1),  37--41 (2019), publisher: LWW

\bibitem{forde-johnston_integrative_2023}
Forde-Johnston, C., Butcher, D., Aveyard, H.: An integrative review exploring the impact of {Electronic} {Health} {Records} ({EHR}) on the quality of nurse-patient interactions and communication. Journal of Advanced Nursing  \textbf{79}(1),  48--67 (Jan 2023)

\bibitem{garber_quantity_2011}
Garber, C.E., Blissmer, B., Deschenes, M.R., Franklin, B.A., Lamonte, M.J., Lee, I.M., Nieman, D.C., Swain, D.P.: Quantity and {Quality} of {Exercise} for {Developing} and {Maintaining} {Cardiorespiratory}, {Musculoskeletal}, and {Neuromotor} {Fitness} in {Apparently} {Healthy} {Adults}: {Guidance} for {Prescribing} {Exercise}. Medicine \& Science in Sports \& Exercise  \textbf{43}(7), ~1334 (Jul 2011). \doi{10.1249/MSS.0b013e318213fefb}, \url{https://journals.lww.com/acsm-msse/fulltext/2011/07000/quantity_and_quality_of_exercise_for_developing.26.aspx}

\bibitem{gartner2025digitally}
Gartner, B., Leysen, D., Mcgowan, H., Wurhofer, D., Eischer, B., Fischer, E., Podolsky, A., Kulnik, S.: Digitally supported shared decision-making for exercise prescription in the secondary prevention of cardiovascular disease. European Journal of Preventive Cardiology  \textbf{32}(Supplement\_1),  zwaf236--362 (2025)

\bibitem{gimpelQuantifyingQuantifiedSelf}
Gimpel, H., Nißen, M.: Quantifying the {Quantified} {Self}: {A} {Study} on the {Motivations} of {Patients} to {Track} {Their} {Own} {Health}

\bibitem{glasgow2014implementation}
Glasgow, R.E., Phillips, S.M., Sanchez, M.A.: Implementation science approaches for integrating {eHealth} research into practice and policy. International journal of medical informatics  \textbf{83}(7),  e1--e11 (2014), publisher: Elsevier

\bibitem{greenhalghAdoptionNewFramework2017}
Greenhalgh, T., Wherton, J., Papoutsi, C., Lynch, J., Hughes, G., A'Court, C., Hinder, S., Fahy, N., Procter, R., Shaw, S.: Beyond {Adoption}: {A} {New} {Framework} for {Theorizing} and {Evaluating} {Nonadoption}, {Abandonment}, and {Challenges} to the {Scale}-{Up}, {Spread}, and {Sustainability} of {Health} and {Care} {Technologies}. Journal of Medical Internet Research  \textbf{19}(11),  e8775 (Nov 2017), company: Journal of Medical Internet Research Distributor: Journal of Medical Internet Research Institution: Journal of Medical Internet Research Label: Journal of Medical Internet Research Publisher: JMIR Publications Inc., Toronto, Canada

\bibitem{haase2023data}
Haase, C.B., Ajjawi, R., Bearman, M., Brodersen, J.B., Risor, T., Hoeyer, K.: Data as symptom: {Doctors}’ responses to patient-provided data in general practice. Social Studies of Science  \textbf{53}(4),  522--544 (2023), publisher: SAGE Publications Sage UK: London, England

\bibitem{InspirationCardWorkshops}
Halskov, K., Dalsg{\aa}rd, P.: Inspiration card workshops. In: Proceedings of the 6th conference on Designing Interactive systems. pp. 2--11 (2006)

\bibitem{haoAdvancingPatientCenteredShared2024}
Hao, Y., Liu, Z., Riter, R.N., Kalantari, S.: Advancing {Patient}-{Centered} {Shared} {Decision}-{Making} with {AI} {Systems} for {Older} {Adult} {Cancer} {Patients}. In: Proceedings of the {CHI} {Conference} on {Human} {Factors} in {Computing} {Systems}. pp. 1--20. {CHI} '24, Association for Computing Machinery, New York, NY, USA (May 2024)

\bibitem{danish2018coherent}
Danish Ministry~of Health, D.M.o.F., Regions, D.: A coherent and trustworthy health network for all. Digital health strategy pp. 2018--2022 (2018)

\bibitem{HttpsWwwGov2014}
of~Health, D., Social~Care, G.o.U.: https://www.gov.uk/government/publications/personalised-health-and-care-2020 (Nov 2014)

\bibitem{heijsters2022stakeholders}
Heijsters, F., Santema, J., Mullender, M., Bouman, M.B., de~Bruijne, M., van Nassau, F.: Stakeholders barriers and facilitators for the implementation of a personalised digital care pathway: a qualitative study. BMJ open  \textbf{12}(11),  e065778 (2022)

\bibitem{herkert_usefulness_2019}
Herkert, C., Kraal, J.J., van Loon, E.M.A., van Hooff, M., Kemps, H.M.C., {others}: Usefulness of modern activity trackers for monitoring exercise behavior in chronic cardiac patients: validation study. JMIR mHealth and uHealth  \textbf{7}(12),  e15045 (2019), publisher: JMIR Publications Inc., Toronto, Canada

\bibitem{giacomini_roadmap_2023}
Hussein, R., Sareban, M., Treff, G., Niebauer, J.: Roadmap for {Aligning} {Cardiovascular} {Digital} {Health} in {Austria} with the {European} {Health} {Data} {Space} ({EHDS}) {Ecosystem}. In: Giacomini, M., Stoicu-Tivadar, L., Balestra, G., Benis, A., Bonacina, S., Bottrighi, A., Deserno, T.M., Gallos, P., Lhotska, L., Marceglia, S., Pazos~Sierra, A.C., Rosati, S., Sacchi, L. (eds.) Studies in {Health} {Technology} and {Informatics}. IOS Press (Oct 2023)

\bibitem{hoppchenBeMeStay2024}
Höppchen, I., Kulnik, S.T., Meschtscherjakov, A., Niebauer, J., Pfannerstill, F., Smeddinck, J.D., Strumegger, E.M., Young, F., Wurhofer, D.: “{Be} with me and stay with me”: {Insights} from {Co}-{Designing} a {Digital} {Companion} to {Support} {Patients} {Transitioning} from {Hospital} to {Cardiac} {Rehabilitation}. In: Proceedings of the 2024 {ACM} {Designing} {Interactive} {Systems} {Conference}. pp. 890--904. {DIS} '24, Association for Computing Machinery, New York, NY, USA (Jul 2024)

\bibitem{hoppchenTargetingBehavioralFactors2024}
Höppchen, I., Wurhofer, D., Meschtscherjakov, A., Smeddinck, J.D., Kulnik, S.T.: Targeting behavioral factors with digital health and shared decision-making to promote cardiac rehabilitation-a narrative review. Frontiers in Digital Health  \textbf{6},  1324544 (2024)

\bibitem{jayathissa_patient-generated_2023}
Jayathissa, P., Sareban, M., Niebauer, J., Hussein, R.: Patient-{Generated} {Health} {Data} {Interoperability} {Through} {Master} {Patient} {Index}: {The} {DH}-{Convener} {Approach}. In: Healthcare {Transformation} with {Informatics} and {Artificial} {Intelligence}, pp. 20--23. IOS Press (2023)

\bibitem{jongsma_how_2021}
Jongsma, K.R., Bekker, M.N., Haitjema, S., Bredenoord, A.L.: How digital health affects the patient-physician relationship: {An} empirical-ethics study into the perspectives and experiences in obstetric care. Pregnancy Hypertension  \textbf{25},  81--86 (Aug 2021)

\bibitem{karanam2014motivational}
Karanam, Y., Filko, L., Kaser, L., Alotaibi, H., Makhsoom, E., Voida, S.: Motivational affordances and personality types in personal informatics. In: Proceedings of the 2014 ACM International Joint Conference on Pervasive and Ubiquitous Computing: Adjunct Publication. pp. 79--82 (2014)

\bibitem{keessenFactorsRelatedFear2020}
Keessen, P., Latour, C.H.M., van Duijvenbode, I.C.D., Visser, B., Proosdij, A., Reen, D., Scholte~op Reimer, W.J.M.: Factors related to fear of movement after acute cardiac hospitalization. BMC Cardiovascular Disorders  \textbf{20}, ~495 (Nov 2020)

\bibitem{khatiwada_patient-generated_2024}
Khatiwada, P., Yang, B., Lin, J.C., Blobel, B.: Patient-{Generated} {Health} {Data} ({PGHD}): {Understanding}, {Requirements}, {Challenges}, and {Existing} {Techniques} for {Data} {Security} and {Privacy}. Journal of Personalized Medicine  \textbf{14}(3), ~282 (Mar 2024). \doi{10.3390/jpm14030282}

\bibitem{kimHowMuchDecision2024}
Kim, D., Vegt, N., Visch, V., Bos-De~Vos, M.: How {Much} {Decision} {Power} {Should} ({A}){I} {Have}?: {Investigating} {Patients}’ {Preferences} {Towards} {AI} {Autonomy} in {Healthcare} {Decision} {Making}. In: Proceedings of the {CHI} {Conference} on {Human} {Factors} in {Computing} {Systems}. pp. 1--17. {CHI} '24, Association for Computing Machinery, New York, NY, USA (May 2024)

\bibitem{kumar_mobile_2021}
Kumar, D., Jeuris, S., Bardram, J.E., Dragoni, N.: Mobile and {Wearable} {Sensing} {Frameworks} for {mHealth} {Studies} and {Applications}: {A} {Systematic} {Review}. ACM Transactions on Computing for Healthcare  \textbf{2}(1),  8:1--8:28 (Dec 2021)

\bibitem{lavalleeMHealthPatientGenerated2020g}
Lavallee, D.C., Lee, J.R., Austin, E., Bloch, R., Lawrence, S.O., McCall, D., Munson, S.A., Nery-Hurwit, M.B., Amtmann, D.: \{{mHealth}\} and patient generated health data: stakeholder perspectives on opportunities and barriers for transforming healthcare. mHealth  \textbf{6}, ~8 (2020)

\bibitem{liStagebasedModelPersonal2010}
Li, I., Dey, A., Forlizzi, J.: A stage-based model of personal informatics systems. In: Proceedings of the {SIGCHI} {Conference} on {Human} {Factors} in {Computing} {Systems}. pp. 557--566. {CHI} '10, Association for Computing Machinery, New York, NY, USA (Apr 2010)

\bibitem{lyonBridgingHCIImplementation2023}
Lyon, A., Munson, S.A., Reddy, M., Schueller, S.M., Agapie, E., Yarosh, S., Dopp, A., von Thiele~Schwarz, U., Doherty, G., Graham, A.K., Kruzan, K.P., Kornfield, R.: Bridging {HCI} and {Implementation} {Science} for {Innovation} {Adoption} and {Public} {Health} {Impact}. In: Extended {Abstracts} of the 2023 {CHI} {Conference} on {Human} {Factors} in {Computing} {Systems}. pp.~1--7. {CHI} {EA} '23, Association for Computing Machinery, New York, NY, USA (Apr 2023)

\bibitem{mastrianni2024ai}
Mastrianni, A., Twede, H., Sarcevic, A., Wander, J., Austin-Tse, C., Saponas, S., Rehm, H., Conard, A.M., Hall, A.K.: Ai-enhanced sensemaking: Exploring the design of a generative ai-based assistant to support genetic professionals. arXiv preprint arXiv:2412.15444  (2024)

\bibitem{mayringQualitativeContentAnalysis2004}
Mayring, P., {others}: Qualitative content analysis. A companion to qualitative research  \textbf{1}(2),  159--176 (2004)

\bibitem{mcgraw2013going}
McGraw, D., Belfort, R., Pfister, H., Ingargiola, S.: Going digital with patients: Managing potential liability risks of patient-generated electronic health information. Journal of Participatory Medicine  \textbf{5} (2013)

\bibitem{medicine_acsms_2013}
Medicine, A.C.o.S., {others}: {ACSM}'s guidelines for exercise testing and prescription. Lippincott williams \& wilkins (2013)

\bibitem{mishraSupportingCollaborativeHealth2018}
Mishra, S.R., {Andrew D. Miller}, {Andrew D. Miller}, Miller, A., Haldar, S., Khelifi, M., Eschler, J., Elera, R.G., Pollack, A.H., Pratt, W.: Supporting {Collaborative} {Health} {Tracking} in the {Hospital}: {Patients}' {Perspectives}  \textbf{2018}, ~650 (Apr 2018), mAG ID: 2795847858

\bibitem{mollerWhoDoesWork2020}
Møller, N.H., Bossen, C., Pine, K.H., Nielsen, T.R., Neff, G.: Who does the work of data? Interactions  \textbf{27}(3),  52--55 (Apr 2020)

\bibitem{nittasElectronicPatientGeneratedHealth2019}
Nittas, V., Lun, P., Ehrler, F., Puhan, M.A., Mütsch, M.: Electronic {Patient}-{Generated} {Health} {Data} to {Facilitate} {Disease} {Prevention} and {Health} {Promotion}: {Scoping} {Review}. Journal of Medical Internet Research  \textbf{21}(10),  e13320 (Oct 2019), company: Journal of Medical Internet Research Distributor: Journal of Medical Internet Research Institution: Journal of Medical Internet Research Label: Journal of Medical Internet Research Publisher: JMIR Publications Inc., Toronto, Canada

\bibitem{nunes2015self}
Nunes, F., Verdezoto, N., Fitzpatrick, G., Kyng, M., Gronvall, E., Storni, C.: Self-care technologies in hci: Trends, tensions, and opportunities. ACM Transactions on Computer-Human Interaction (TOCHI)  \textbf{22}(6),  1--45 (2015)

\bibitem{oh_patients_2022}
Oh, C.Y., Luo, Y., St.~Jean, B., Choe, E.K.: Patients {Waiting} for {Cues}: {Information} {Asymmetries} and {Challenges} in {Sharing} {Patient}-{Generated} {Data} in the {Clinic}. Proc. ACM Hum.-Comput. Interact.  \textbf{6}(CSCW1),  107:1--107:23 (Apr 2022)

\bibitem{noauthor_cardiovascular_nodate}
Organization, W.H.: Cardiovascular diseases ({CVDs}) (2021)

\bibitem{pine2018data}
Pine, K.H., Bossen, C., Chen, Y., Ellingsen, G., Grisot, M., Mazmanian, M., M{\o}ller, N.H.: Data work in healthcare: Challenges for patients, clinicians and administrators. In: Companion of the 2018 ACM Conference on Computer Supported Cooperative Work and Social Computing. pp. 433--439 (2018)

\bibitem{rajUnderstandingIndividualCollaborative2017b}
Raj, S., Newman, M.W., Lee, J.M., Ackerman, M.S.: Understanding {Individual} and {Collaborative} {Problem}-{Solving} with {Patient}-{Generated} {Data}: {Challenges} and {Opportunities}. Proc. ACM Hum.-Comput. Interact.  \textbf{1}(CSCW),  88:1--88:18 (Dec 2017)

\bibitem{radar2019}
Ranjan, Y., Rashid, Z., Stewart, C., Conde, P., Begale, M., Verbeeck, D., Boettcher, S., Dobson, R., Folarin, A.: Radar-base: Open source mobile health platform for collecting, monitoring, and analyzing data using sensors, wearables, and mobile devices. JMIR Mhealth Uhealth  \textbf{7}(8),  e11734 (Aug 2019)

\bibitem{rosman_when_2020}
Rosman, L., Gehi, A., Lampert, R.: When smartwatches contribute to health anxiety in patients with atrial fibrillation. Cardiovascular Digital Health Journal  \textbf{1}(1),  9--10 (Aug 2020)

\bibitem{royCardbasedDesignTools2019}
Roy, R., Warren, J.P.: Card-based design tools: a review and analysis of 155 card decks for designers and designing. Design Studies  \textbf{63},  125--154 (Jul 2019)

\bibitem{rumboldWhatAreData2018a}
Rumbold, J.M.M., Pierscionek, B.K.: What {Are} {Data}? {A} {Categorization} of the {Data} {Sensitivity} {Spectrum}. Big Data Research  \textbf{12},  49--59 (Jul 2018)

\bibitem{rutjesBenefitsCostsPatient2017}
Rutjes, H., Willemsen, M.C., van Kollenburg, J., Bogers, S., IJsselsteijn, W.A.: Benefits and costs of patient generated data, from the clinician's and patient's perspective. In: Proceedings of the 11th {EAI} {International} {Conference} on {Pervasive} {Computing} {Technologies} for {Healthcare}. pp. 436--439. {PervasiveHealth} '17, Association for Computing Machinery, New York, NY, USA (May 2017)

\bibitem{sandersGenerativeToolsCodesigning2000}
Sanders, E.B.N.: Generative {Tools} for {Co}-designing. In: Scrivener, S.A.R., Ball, L.J., Woodcock, A. (eds.) Collaborative {Design}. pp. 3--12. Springer, London (2000)

\bibitem{sareban2025op}
Sareban, M., Treff, G., Smeddinck, J.D., Hussein, R., Niebauer, J.: Opportunities and barriers for reimbursement of digital therapeutics in austria: Findings from expert interviews. DIGITAL HEALTH  \textbf{11},  20552076241299062 (2025)

\bibitem{scholich_augmenting_2024}
Scholich, T., Raj, S., Lee, J., Newman, M.W.: Augmenting clinicians’ analytical workflow through task-based integration of data visualizations and algorithmic insights: a user-centered design study. Journal of the American Medical Informatics Association  \textbf{31}(11),  2455--2473 (Nov 2024)

\bibitem{schroeder2017supporting}
Schroeder, J., Hoffswell, J., Chung, C.F., Fogarty, J., Munson, S., Zia, J.: Supporting patient-provider collaboration to identify individual triggers using food and symptom journals. In: Proceedings of the 2017 ACM conference on computer supported cooperative work and social computing. pp. 1726--1739 (2017)

\bibitem{serbanJustSeeNumbers2023}
Serban, I.B., Dritsa, D., Campero~Jurado, I., Houben, S., Brombacher, A., Ten~Cate, D., Janssen, L., Heijmans, M.: “{I} just see numbers, but how do you feel about your training?”: {Clinicians}' {Data} {Needs} in {Telemonitoring} for {Colorectal} {Cancer} {Surgery} {Prehabilitation}. In: Companion {Publication} of the 2023 {Conference} on {Computer} {Supported} {Cooperative} {Work} and {Social} {Computing}. pp. 267--272. {CSCW} '23 {Companion}, Association for Computing Machinery, New York, NY, USA (Oct 2023)

\bibitem{shneidermanEyesHaveIt1996}
Shneiderman, B.: The eyes have it: a task by data type taxonomy for information visualizations. In: Proceedings 1996 {IEEE} {Symposium} on {Visual} {Languages}. pp. 336--343 (Sep 1996), iSSN: 1049-2615

\bibitem{sinsky_allocation_2016}
Sinsky, C., Colligan, L., Li, L., Prgomet, M., Reynolds, S., Goeders, L., Westbrook, J., Tutty, M., Blike, G.: Allocation of {Physician} {Time} in {Ambulatory} {Practice}: {A} {Time} and {Motion} {Study} in 4 {Specialties}. Annals of Internal Medicine  \textbf{165}(11),  753--760 (Dec 2016)

\bibitem{sittigNewSociotechnicalModel2010}
Sittig, D.F., Singh, H.: A {New} {Socio}-technical {Model} for {Studying} {Health} {Information} {Technology} in {Complex} {Adaptive} {Healthcare} {Systems}. Quality \& safety in health care  \textbf{19}(Suppl 3),  i68--i74 (Oct 2010)

\bibitem{stromelNarratingFitnessLeveraging2024}
Strömel, K.R., Henry, S., Johansson, T., Niess, J., Woźniak, P.W.: Narrating {Fitness}: {Leveraging} {Large} {Language} {Models} for {Reflective} {Fitness} {Tracker} {Data} {Interpretation}. In: Proceedings of the {CHI} {Conference} on {Human} {Factors} in {Computing} {Systems}. pp. 1--16. {CHI} '24, Association for Computing Machinery, New York, NY, USA (May 2024)

\bibitem{suWhatYourEnvisioned2022}
Su, Z., He, L., Jariwala, S.P., Zheng, K., Chen, Y.: "{What} is {Your} {Envisioned} {Future}?": {Toward} {Human}-{AI} {Enrichment} in {Data} {Work} of {Asthma} {Care}. Proc. ACM Hum.-Comput. Interact.  \textbf{6}(CSCW2),  267:1--267:28 (Nov 2022)

\bibitem{tadas_user-centred_2022}
Tadas, S.: User-centred {Digital} {Health} in {Cardiovascular} {Rehabilitation} and {Self}-management. {PhD} {Thesis}, University College Dublin. School of Computer Science (2022)

\bibitem{tadasUsingPatientGeneratedData2023a}
Tadas, S., Dickson, J., Coyle, D.: Using {Patient}-{Generated} {Data} to {Support} {Cardiac} {Rehabilitation} and the {Transition} to {Self}-{Care}. In: Proceedings of the 2023 {CHI} {Conference} on {Human} {Factors} in {Computing} {Systems}. pp. 1--16. {CHI} '23, Association for Computing Machinery, New York, NY, USA (Apr 2023)

\bibitem{tawfik_physician_2018}
Tawfik, D.S., Profit, J., Morgenthaler, T.I., Satele, D.V., Sinsky, C.A., Dyrbye, L.N., Tutty, M.A., West, C.P., Shanafelt, T.D.: Physician {Burnout}, {Well}-being, and {Work} {Unit} {Safety} {Grades} in {Relationship} to {Reported} {Medical} {Errors}. Mayo Clinic Proceedings  \textbf{93}(11),  1571--1580 (Nov 2018)

\bibitem{topolskiPeerReviewedRapid2006}
Topolski, T.D., LoGerfo, J., Patrick, D.L., Williams, B., Walwick, J., Patrick, M.M.B.: Peer reviewed: the {Rapid} {Assessment} of {Physical} {Activity} ({RAPA}) among older adults. Preventing chronic disease  \textbf{3}(4) (2006), publisher: Centers for Disease Control and Prevention

\bibitem{VSPakianathan2024}
V~S~Pakianathan, P., Kumar, D., Jayatissa, P., Rada, H., Niebauer, J., Schmidt, A., Smeddinck, J.: Barriers and enablers in integrating patient-generated health data for shared decision-making between healthcare professionals and patients: A scoping review (preprint)  (May 2024)

\bibitem{v_s_pakianathan_multi-stakeholder_2023}
V~S~Pakianathan, P., Wurhofer, D., Kumar, D., Niebauer, J., Smeddinck, J.D.: Multi-{Stakeholder} {Design} for {Complex} {Digital} {Health} {Systems}: {Development} of a {Modular} {Open} {Research} {Platform} ({MORE}). In: {dHealth}. IOS Press (2023)

\bibitem{varma2024promises}
Varma, N., Han, J.K., Passman, R., Rosman, L.A., Ghanbari, H., Noseworthy, P., Avari~Silva, J.N., Deshmukh, A., Sanders, P., Hindricks, G., et~al.: Promises and perils of consumer mobile technologies in cardiovascular care: Jacc scientific statement. Journal of the American College of Cardiology  \textbf{83}(5),  611--631 (2024)

\bibitem{waddellLeveragingImplementationScience2024a}
Waddell, A., Seguin, J.P., Wu, L., Stragalinos, P., Wherton, J., Watterson, J.L., Prawira, C.O., Olivier, P., Manning, V., Lubman, D., Grigg, J.: Leveraging {Implementation} {Science} in {Human}-{Centred} {Design} for {Digital} {Health}. In: Proceedings of the {CHI} {Conference} on {Human} {Factors} in {Computing} {Systems}. pp. 1--17. {CHI} '24, Association for Computing Machinery, New York, NY, USA (May 2024)

\bibitem{west2018common}
West, P., Van~Kleek, M., Giordano, R., Weal, M.J., Shadbolt, N.: Common barriers to the use of patient-generated data across clinical settings. In: proceedings of the 2018 CHI Conference on Human Factors in Computing Systems. pp. 1--13 (2018)

\bibitem{winnige2021cardiac}
Winnige, P., Vysoky, R., Dosbaba, F., Batalik, L.: Cardiac rehabilitation and its essential role in the secondary prevention of cardiovascular diseases. World journal of clinical cases  \textbf{9}(8), ~1761 (2021), publisher: Baishideng Publishing Group Inc

\bibitem{wurhoferInvestigatingSharedDecisionmaking2024}
Wurhofer, D., Neunteufel, J., Strumegger, E.M., Höppchen, I., Mayr, B., Egger, A., Sareban, M., Reich, B., Neudorfer, M., Niebauer, J., Smeddinck, J.D., Kulnik, S.T.: Investigating shared decision-making during the use of a digital health tool for physical activity planning in cardiac rehabilitation. Frontiers in Digital Health  \textbf{5} (Jan 2024), publisher: Frontiers

\bibitem{yeImpactElectronicHealth2021a}
Ye, J.: The impact of electronic health record–integrated patient-generated health data on clinician burnout. Journal of the American Medical Informatics Association  \textbf{28}(5),  1051--1056 (May 2021)

\bibitem{yiminlimDatastormingCraftingData2021}
Yi~Min~Lim, D., Yap, C.E.L., Lee, J.J.: Datastorming: {Crafting} {Data} into {Design} {Materials} for {Design} {Students}’ {Creative} {Data} {Literacy}. In: Proceedings of the 13th {Conference} on {Creativity} and {Cognition}. pp.~1--9. C\&amp;{C} '21, Association for Computing Machinery, New York, NY, USA (Jun 2021)

\bibitem{zhang2021Data}
Zhang, Z., Joy, K., Upadhyayula, P., Ozkaynak, M., Harris, R., Adelgais, K.: Data work and decision making in emergency medical services: A distributed cognition perspective. Proc. ACM Hum.-Comput. Interact.  \textbf{5}(CSCW2) (Oct 2021)

\bibitem{zhu_sharing_2016}
Zhu, H., {Joanna Colgan}, Colgan, J., Reddy, M.C., Choe, E.K.: Sharing {Patient}-{Generated} {Data} in {Clinical} {Practices}: {An} {Interview} {Study}. American Medical Informatics Association Annual Symposium  \textbf{2016},  1303--1312 (Jan 2016), mAG ID: 2621378336 S2ID: 37af029ee6789753e1a088b79030c4aaf029b54a

\end{thebibliography}

\end{document}